\documentclass[fleqn,usenatbib]{mnras}

\usepackage{newtxtext,newtxmath}

\usepackage[T1]{fontenc}
\usepackage[pinyin,australian]{babel}
\DeclareRobustCommand{\VAN}[3]{#2}
\let\VANthebibliography\thebibliography
\def\thebibliography{\DeclareRobustCommand{\VAN}[3]{##3}\VANthebibliography}

\usepackage{graphicx}	
\usepackage{amsmath}	

\title[Effective2-body Scattering]{Effective two-body scatterings around a massive object}

\author[Y. Wang et al.]{
Yihan Wang,$^{1,2}$\thanks{E-mail: yihan.wang@unlv.edu}
Saavik Ford,$^{4,5}$
Rosalba Perna,$^{3,5}$
Barry McKernan,$^{4,5}$
Zhaohuan Zhu$^{1,2}$
and Bing Zhang$^{1,2}$\\
$^{1}$Nevada Center for Astrophysics, University of Nevada, Las Vegas, NV 89154\\
$^{2}$Department of Physics and Astronomy, University of Nevada Las Vegas, Las Vegas, NV 89154, USA\\
$^{3}$Department of Physics and Astronomy, Stony Brook University, Stony Brook, NY 11794-3800, USA\\
$^{4}$Department of Science, BMCC, City University of New York, New York, NY 10007, USA\\
$^{5}$Center for Computational Astrophysics, Flatiron Institute, New York, NY 10010, USA\\
}

\date{Accepted XXX. Received YYY; in original form ZZZ}

\pubyear{2015}

\begin{document}
\label{firstpage}
\pagerange{\pageref{firstpage}--\pageref{lastpage}}
\maketitle

\begin{abstract}
Two-body scatterings under the potential of a massive object are very common in astrophysics. If the massive body is far enough away that the two small bodies are in their own gravitational sphere of influence, the gravity of the massive body can be temporarily ignored. However, this requires the scattering process to be fast enough that the small objects do not spend too much time at distances near the surface of the sphere of influence. In this paper, we derive the validation criteria for effective two-body scattering and establish a simple analytical solution for this  process, which we verify through numerical scattering experiments.  
We use this solution to study star-black hole scatterings in the disks of Active Galactic Nuclei and planet-planet scatterings in planetary systems, and calculate their one-dimensional cross-section analytically. Our solution will be valuable in reducing computational time when treating two-body scatterings under the potential of a much more massive third body, provided that the problem settings are in the valid parameter space region identified by our study.
\end{abstract}


\section{Introduction}

The study of two-body scatterings is a fundamental topic in physics that has been investigated for over a century. When two particles interact, they exchange energy and momentum, resulting in a change in their direction and speed. When such scattering events occur around a massive object, the particles can experience a gravitational deflection due to the object's gravitational field, altering their impact parameter and scattering angle.

The concept of the gravitational sphere of influence was first introduced by Laplace to study close encounters between comets and Jupiter in our solar system. This sphere defines the region where the motion of objects is dominated by the gravity of a celestial body, and the gravity of other objects can be ignored temporarily. The radius of the sphere can be defined in several ways, such as Hill's radius and Laplace's radius, which depend solely on the masses of the objects involved.

 Öpik \citep{Opik1976} introduced a model for calculating the trajectories of asteroids following close encounters with planets. The model assumes encounters are instantaneous and that the gravity of the Sun can be temporarily ignored, which allows for the explicit expression of the post-encounter orbital parameters of the asteroid from two-body scattering with pre-encounter initial conditions. If the encounter is fast enough that the orbital deflection from the planet is negligible, Opik's method works well over a wide range of parameter space. However, when orbital deflection from the planet is significant, this method fails to accurately predict the post-encounter trajectories of the scattering objects.

Initially, Opik's method was thought to be useful only when the impact parameter between the asteroid and the planet was smaller than the radius of the sphere of influence, and the two-body scattering model became invalid when the impact parameter was too large. However, later investigations \citep{1988Icar...75....1G,1990CeMDA..49..111C,1997P&SS...45.1561V} found that Opik's method becomes less reliable because the pre-encounter deflection from the planet results in a different encounter geometry than originally assumed. However, if this deflection can be accurately calculated or a more solid velocity-dependent (implicitly or explicitly) sphere of influence can be used to validate the parameter space of Opik's method, accurate analytical post-encounter trajectory predictions are still possible. The goal of this paper is to identify the accurate parameter space in which Opik's method can be properly used and apply the method to several astrophysical phenomena to obtain analytical cross-sections for these events. 

Fortunately, in many cases of astrophysical interest, we are in a regime that does permit analytic solutions that are valid to high accuracy. In particular, in the case of a star with planets or protoplanets in a disk, or the analogous situation of stars or stellar mass black holes in an active galactic nucleus (AGN) disk, the gas disk provides a preferred orientation for orbits around the central object. 

Stars and stellar-origin BHs are expected to be found in the disks of AGNs, either due to in-situ formation \citep[e.g.][]{Stone2017} or due to capture from the nuclear star clusters \citep[e.g.][]{McKernan12,Kennedy16,Bartos2017,Fabj2020}.
Once in the disk, BHs and stars are subject to frequent dynamical interactions \citep{Samsing2020,Wang2021AGN}, and our formalism allows us to easily identify regions of parameter space in which the outcome of the scattering is especially interesting. In particular, in the following we will consider in detail two cases: one in which the scattering leads to the tidal disruption of the star by the stellar mass BH (micro-TDE) \citep{Perets2016,Kremer2022,Yang20,Wang2021,Ryu2022}, and one in which the star gets scattered within the sphere of influence (for tidal disruption) of the central SMBH, hence giving rise to a standard tidal disruption event (TDE)  \citep{Rees88,Evans89,Phinney89}.

The paper is structured as follows: in section 2 we describe the problem and the analytical solutions for the post-scattering trajectories, and present analytical formulas suitable for inclusion into $N$-body codes; we further test these formulas against numerical simulations.
In section 3 we apply our results to 
two astrophysical scenarios:  
an AGN disk, for which we determine the rate of TDEs and micro-TDEs, and
a protoplanetary/transition disk, where we determine the distribution of free-floating
and highly eccentric planets. We finally 
summarize and discuss the caveats of our analysis in Sec.4.

\section{Hyperbolic scatterings around a massive object}

\subsection{Free two-body scattering and turning angle}

To construct our formalism, we begin with free two-body scattering and we consider the turning angle after the encounter. The scattering object is assumed to be on a hyperbolic orbit with semi-major axis $a_{\rm hyp}$, eccentricity $e_{\rm hyp}$, and total energy of the two-body system $> 0$. If we choose one of the objects as a reference frame, the distance of the other object to the origin can be written as
\begin{equation}
    r_{\rm hyp} = \frac{P}{1+e_{\rm hyp}\cos\theta}\,,\label{eq:traj}
\end{equation}
where $P=a_{\rm hyp}(1-e_{\rm hyp}^2)$ is the semi-latus rectum, and $\theta$ is the true anomaly. $\theta=0$ yields $r=a_{\rm hyp}(1-e_{\rm hyp})$, which corresponds to the pericenter distance. Using energy conservation we can calculate the semi-major axis of the hyperbolic orbit. The specific total energy of the system at $t=-\infty$ is ${v_\infty^2}/{2}$, while the total specific energy of the Keplerian orbit is $-\mu_{12}/{2a_{\rm hyp}}$, where $\mu_{ijk...}=G(m_i+m_j+m_k+...)$ is the gravitational constant of an $N$-body system. Therefore, from $\frac{v_\infty^2}{2}=-\frac{\mu_{12}}{2a_{\rm hyp}}$, we get
\begin{equation}
    a_{\rm hyp} = -\frac{\mu_{12}}{v_\infty^2}\,.
\end{equation}
The eccentricity of a conic section is given by
\begin{equation}
    e_{\rm hyp} = \sqrt{1+b^2/a_{\rm hyp}^2}\,,
\end{equation}
where $b$ is the impact parameter.
Using Equation~\ref{eq:traj}, we can calculate the corresponding $\theta(t)$ of $r_{\rm hyp}=+\infty$,
\begin{equation}
    \theta(\pm\infty) = \pm\arccos(1/e_{\rm hyp})\,.
\end{equation}
The turning angle of the hyperbolic trajectory, $\theta_{\rm turn}=\pi- [\theta(+\infty) - \theta(-\infty)]$  can then be written as 
\begin{equation}
    \theta_{\rm turn} = \pi-2 \arccos \left(\frac{1}{e_{\rm hyp}}\right)\,.
\end{equation}
Plugging in the semi-major axis and eccentricity $a_{\rm hyp}$ and $e_{\rm hyp}$, we can finally express the turning angle as a function of $v_\infty$ and $b$,
\begin{equation}
     \theta_{\rm turn} = 2 \arctan\frac{\mu_{12}}{v_\infty^2b}\,.
\end{equation}

\subsection{Turning timescale}
The time around the closest approach $\tau_{\rm turn} = t(\theta=\pi/2) - t(\theta=-\pi/2)$ can be calculated as
\begin{equation}
 \Delta t = \sqrt{-a_{\rm hyp}^3/\mu_{12}} [M(t_{\rm f}) - M(t_{\rm 0})]\,,
\end{equation}
where $M(t)$ is the mean anomaly (the fraction of a Keperian orbit's period that has elapsed) of the hyperbolic orbit. This corresponds to the time that the relative distance between $m_1$ and $m_2$ is smaller than $a_{\rm hyp}(1-e^2_{\rm hpy})$ (Later on, we will show that this is consistent with the well-known Hill's sphere of influence and our experiments find that this is a good approximation for the turning time). For $\theta = \pm\pi/2$, the corresponding $M(t)$ are 

\begin{eqnarray}
M_{\theta=\pi/2} &=& e_{\rm hyp}\sqrt{e_{\rm hyp}^2-1}-\ln(e_{\rm hyp}+\sqrt{e_{\rm hyp}^2-1})\nonumber\\
&=&f(e_{\rm hyp})\\
M_{\theta=-\pi/2} &=& -e_{\rm hyp}\sqrt{e_{\rm hyp}^2-1}+\ln(e_{\rm hyp}+\sqrt{e_{\rm hyp}^2-1})\,,
\end{eqnarray}
respectively. Therefore, the timescale of the scattering event (timescale to turn) can be estimated as
\begin{equation}
    \tau_{\rm turn} = 2f(e_{\rm hyp}) \sqrt{-a_{\rm hyp}^3/\mu_{12}}=2\sqrt{\frac{P^3}{\mu_{12}}}\frac{f(e_{\rm hyp})}{(e_{\rm hyp}^2-1)^{3/2}}=2\sqrt{\frac{P^3}{\mu_{12}}}g(e_{\rm hyp})\label{eq:t-sac}\,.
\end{equation}
The semi-latus rectum $P$ can also be rewritten as $P=h^2/\mu_{12}=b^2v_\infty^2/\mu_{12}$, where $h$ is the specific angular momentum.

\subsection{Effective two-body scattering in the potential of a third massive object}

If the two-body scattering happens under the presence of a massive object, at the closest approach between the two light objects, the center of mass of these will undergo a nearly Keplerian motion around the massive object. The corresponding timescale of this motion is 
\begin{equation}
    \tau_{\rm orb} = 2\pi\sqrt{r^3/\mu_{123}}\sim 2\pi\sqrt{r^3/\mu_{3}}\,,
\end{equation}
where $m_1$ and $m_2$ are the masses of the light objects and $m_3$ the mass of the heavier one. Comparing this timescale with the turning timescale we obtained in the last subsection, we obtain the timescale ratio
\begin{equation}
   \tau_{\rm turn}/\tau_{\rm orb} =\frac{g(e_{\rm hyp})}{\pi}\left(\frac{\mu_3}{\mu_{12}}\right)^{1/2}\left(\frac{bv_\infty}{\sqrt{\mu_{12}r}}\right)^3\,.
\end{equation}
If the ratio $\tau_{\rm turn}/\tau_{\rm orb}$ is small enough, the center of mass movements around the massive object can be safely ignored during the scattering process. 

\begin{figure}
\includegraphics[width=\columnwidth]{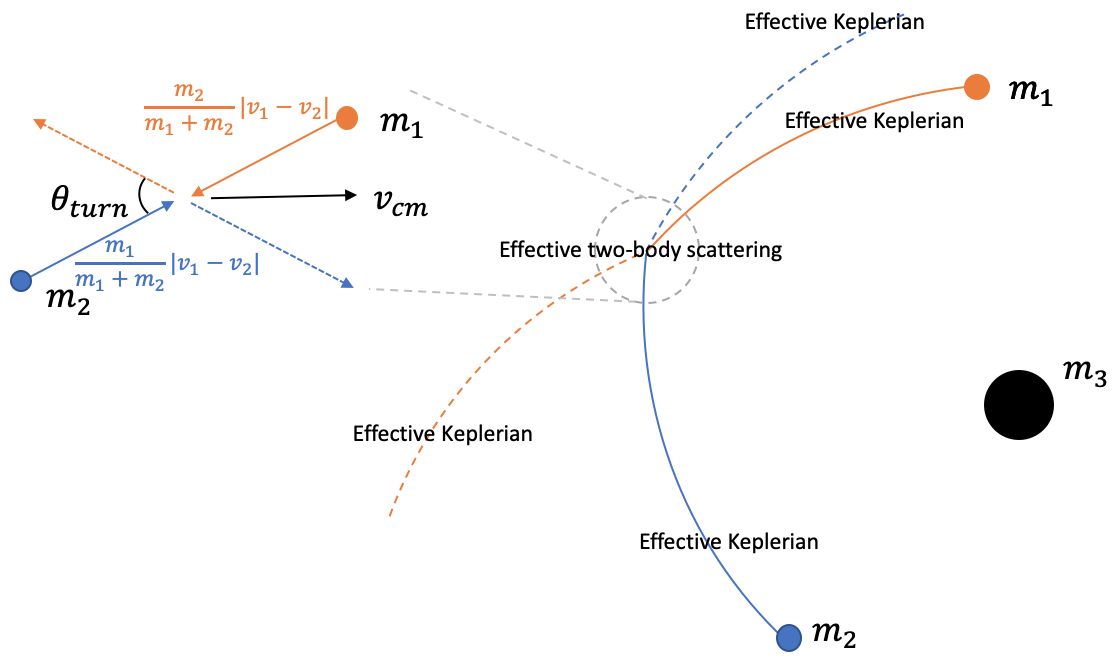}
\includegraphics[width=\columnwidth]{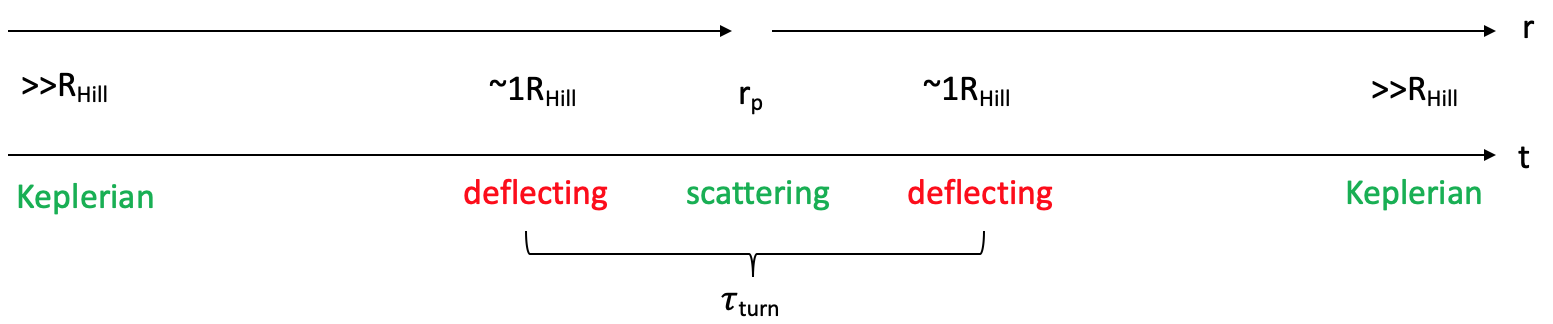}
\caption{ Schematics of effective two-body scattering theory. If $\tau_{\rm turn}/\tau_{\rm orb}\ll 1$, the scattering keeps the center of mass velocity $\mathbf{v}_{\rm cm}$ but turns the reduced velocity vectors by an angle of $\theta_{\rm turn}$.}
\label{fig:2body-schematics}
\end{figure}

\begin{figure}

\includegraphics[width=\columnwidth]{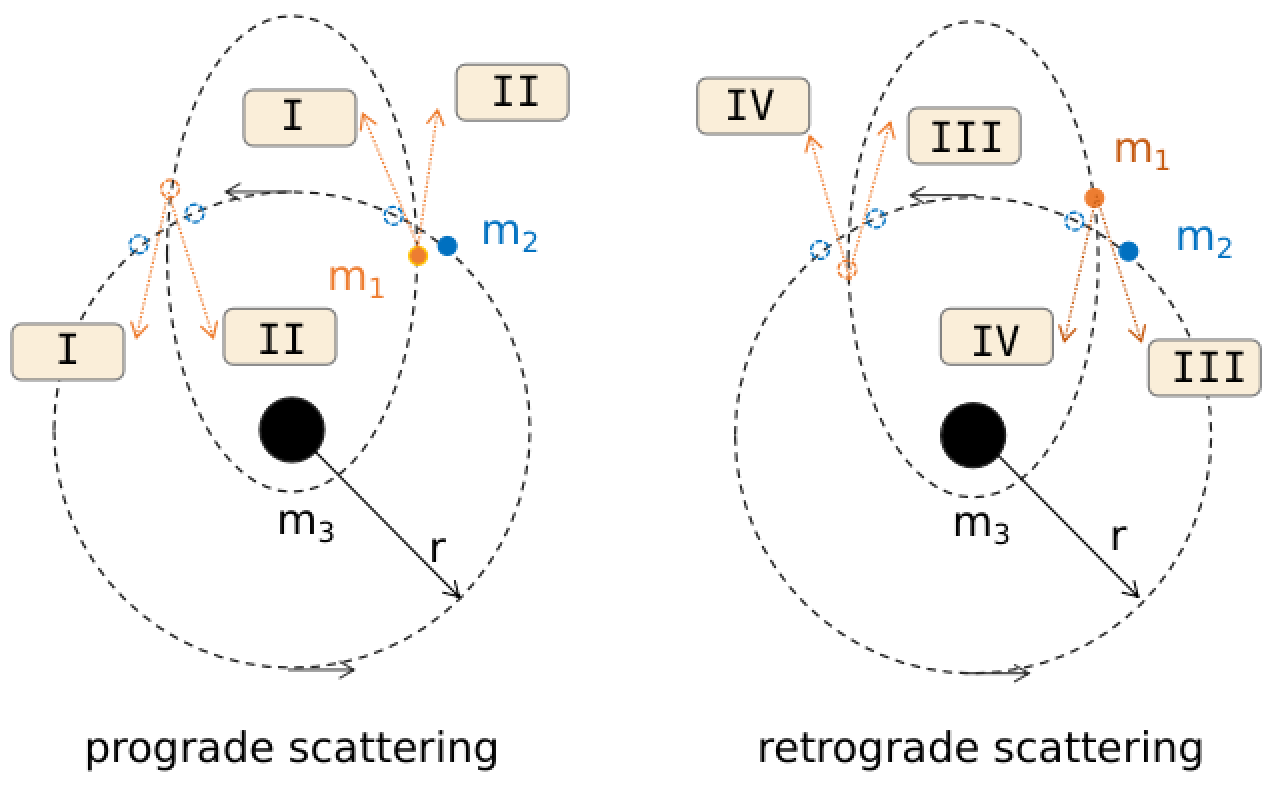}
\caption{Schematics of disk-like two-body scatterings around massive object $m_3$ in Keplerian orbits. \textit{Left}: panel shows the prograde scatterings where $m_1$ and $m_2$ orbit around $m_3$ in the same (\textbf{counter-clockwise}) direction. \textit{Right}: panels shows the retrograde scatterings where $m_2$ orbits as before, but $m_1$ now orbits around $m_3$ in the opposite (\textbf{clockwise}) direction.}
\label{fig:schematics}
\end{figure}

In the reference frame of the third massive body $m_3$, before the scattering, if the velocity of $m_1$ is $\mathbf{v}_1$ and the velocity of  $m_2$ is $\mathbf{v}_2$, we can rewrite these velocities as
\begin{eqnarray}
\mathbf{v}_1 = \frac{m_1\mathbf{v}_1+m_2\mathbf{v}_2}{m_{12}} + \frac{m_2}{m_{12}}(\mathbf{v}_1-\mathbf{v}_2)\label{eq:v1}\\
\mathbf{v}_2 = \frac{m_1\mathbf{v}_1+m_2\mathbf{v}_2}{m_{12}} + \frac{m_1}{m_{12}}(\mathbf{v}_2-\mathbf{v}_1)\label{eq:v2}\,.
\end{eqnarray}
The first term in each equation is the center of mass velocity of $m_1$ and $m_2$, while the second term represents the velocity in their centre of mass reference frame. If the condition \begin{equation}
    \tau_{\rm turn}/\tau_{\rm orb} \ll 1
\end{equation}
is satisfied, the motion of the center of mass of $m_1$ and $m_2$ can be safely ignored during the turning time. Therefore, during this time, only the second term of Equation~\ref{eq:v1} and \ref{eq:v2} changes. As shown in Figure~\ref{fig:2body-schematics}, in the centre of mass reference frame of $m_1$ and $m_2$, due to energy and angular momentum conservation, the scattering maintains the magnitude of the velocities of $m_1$ and $m_2$ but turns them by an angle of $\theta_{\rm turn}$. Therefore, in the reference frame of $m_3$, the velocity of $m_1$ and $m_2$ after the scattering can be expressed as
\begin{eqnarray}
    \mathbf{v}_1^\prime = \frac{m_1\mathbf{v}_1+m_2\mathbf{v}_2}{m_{12}} + \frac{m_2}{m_{12}}(\mathbf{v}_1-\mathbf{v}_2)\mathscr{R}(\theta_{\rm turn})\label{eq:v1p}\\
\mathbf{v}_2^\prime = \frac{m_1\mathbf{v}_1+m_2\mathbf{v}_2}{m_{12}} + \frac{m_1}{m_{12}}(\mathbf{v}_2-\mathbf{v}_1)\mathscr{R}(\theta_{\rm turn})\label{eq:v2p}
\end{eqnarray}
where $\mathscr{R}(\theta_{\rm turn})$ is the rotation matrix on scattering plane, with $\theta_{\rm turn}\sim2 \arctan\frac{\mu_{12}}{|\mathbf{v}_1-\mathbf{v}_2|^2b}$. The position vectors of $m_1$ and $m_2$ can be assumed to be unchanged (the scattering region is very small compared to $r$).

\subsection{Scattering between two equal energy Keplerian orbits}
The simplest scattering configuration between $m_1$ and $m_2$ in the potential of $m_3$ is one in which $m_1$ and $m_2$ move in Keperian orbits and encounter at a location with distance $r$ from $m_3$. In this situation, the corresponding velocities and positions of $m_1$ and $m_2$ in the $m_3$ reference frame are 
\begin{eqnarray}
\mathbf{p}_1 &=& r(\cos\nu_1,\sin\nu_1,0)\mathscr{E}(\omega,i,\Omega)\\
\mathbf{p}_2 &=& r(\cos\nu_2,\sin\nu_2,0)\mathscr{E}(\omega,i,\Omega)\\
\mathbf{v}_1 &=& \pm\sqrt{\frac{\mu_{13}}{p_1}}(-\sin\nu_1,e_1+\cos\nu_1,0)\mathscr{E}(\omega,i,\Omega)\\
\mathbf{v}_2 &=& \pm\sqrt{\frac{\mu_{23}}{p_2}}(-\sin\nu_2,e_2+\cos\nu_2,0)\mathscr{E}(\omega,i,\Omega)
\end{eqnarray}
where $\mu_{j3} = G(m_j+M_3)$, $p_j$ and $e_j$ are the semi-latus rectum  and eccentricity of $m_j$'s orbit, and $\nu_1=\nu_2$ is the true anomaly that is obtained via $\cos\nu_j = (p_j/r-1)/e_j$. $\mathscr{E}(\omega,i,\Omega)$ is the Euler rotation matrix with argument periapsis $\omega$, inclination $i$ and longitude of the ascending node $\Omega$. 

Scatterings between an object ($m_2$) in a circular orbit and one in an eccentric orbit ($m_1$) with the same semi-major axis $a_1=a_2=r$ can be used to demonstrate most of the possible outcomes of the two-body coplanar scattering around a massive body.  As shown in Figure~\ref{fig:schematics}, there are eight  possible scatterings in this scenario due to different orbiting directions and relative positions of $m_1$ and $m_2$. However, because of the symmetry of the orbits, there are scatterings that result in the same post-scattered orbits. Thus, only four unique scatterings exist. We label the symmetric scatterings with the same number labels as in Figure~\ref{fig:schematics}.

\subsection{Valid parameter space}
To safely ignore the gravity of $m_3$ during the scattering process between $m_1$ and $m_2$, the timescale ratio $ \tau_{\rm turn}/\tau_{\rm orb}$ needs to be much smaller than one, 
\begin{eqnarray}
    \tau_{\rm turn}/\tau_{\rm orb}=\frac{1}{\pi}g\bigg(\sqrt{1+\frac{b^2v_\infty^4}{\mu_{12}^2}}\bigg)\left(\frac{\mu_3}{\mu_{12}}\right)^{1/2}\left(\frac{bv_\infty}{\sqrt{\mu_{12}r}}\right)^3 \ll 1\,.
\end{eqnarray}
Let us first consider the case of scattering between two objects in circular orbits; we then  derive\footnote{The Taylor expansion of g($e_{\rm hyp}$) at $e_{\rm hyp}$=1 and $e_{\rm hyp}$=$\infty$ are g($e_{\rm hyp}$)$\sim \frac{2}{3}-\frac{2(e_{\rm hyp}-1)}{5}+O((e_{\rm hyp}-1)^2)$ and g($e_{\rm hyp}$)$\sim \frac{1}{e_{\rm hyp}}+\frac{1-\log(2e_{\rm hyp})}{e_{\rm hyp}^3}+O(\frac{1}{e_{\rm hyp}^5})$, respectively. For prograde scatterings, $v_\infty$ is $b/r$ times of orbital velocity at $r$ and $e_{\rm hyp}\rightarrow1$. For retrograde scatterings, $v_\infty$ is two times of orbital velocity at $r$ and $e_{\rm hyp}$ is very large that g($e_{\rm hyp}$) can be effectively expanded at infinity.}
\begin{eqnarray}\label{eq:valid}
b \ll
\left\{
\begin{aligned}
&\bigg(\frac{9\pi^3}{8}\bigg)^{1/6}\bigg(\frac{m_{12}}{m_{3}}\bigg)^{1/6}R_{\rm Hill}\sim 1.8\bigg(\frac{m_{12}}{m_{3}}\bigg)^{1/6}R_{\rm Hill},\quad {\rm retrograde}\\
&\bigg(\frac{27\pi}{16}\bigg)^{1/6}R_{\rm Hill}\sim 1.3R_{\rm Hill},\quad {\rm prograde}
\end{aligned}
\right.
\end{eqnarray}
The prograde case is the well-known Hill's radius that has been obtained in the same prograde configuration by balancing the gravity between the small objects $m_1$, $m_2$, and the massive object $m_3$. For the retrograde case, where the relative velocity between $m_1$ and $m_2$ is larger than for the prograde case, one might expect that the scattering is faster due to the larger relative velocity. However,  $a_{\rm hyp}$ is much larger than the one in the retrograde case. It takes a longer time for $m_1$ to fly out of the region $r<a_{\rm hyp}(1-e^2_{\rm hpy})$ to be an asymptotic straight line. Therefore, a smaller impact parameter is required for a two-body scattering approximation. Because these two cases give the extreme values of $v_\infty$ (retrograde case gives the maximum value while prograde case gives the minimum value), for other scatterings with non-zero eccentricities, the critical impact parameter $b$ is bracketed in between these two extreme values.

One should note that our numerical scattering experiments indicate that in the extreme case of prograde circular scattering with zero eccentricity, $m_1$ and $m_2$ may undergo continuous resonance scatterings so that no complete single scattering can be found in the continuous scattering patterns. The effective two-body scattering model in this paper can only be used for single scatterings in which $m_1$ and $m_2$ become unbound after the first encounter.

\subsection{Post-scattering orbital calculations}
The post-scattered velocities of $m_1$ and $m_2$ can be obtained by using Equation~\ref{eq:v1p} to \ref{eq:v2p}. Then the orbital parameters of $m_1$ and $m_2$ can be calculated by
\begin{eqnarray}
a_i &=& \frac{\mu_{i3}}{\frac{2\mu_{i3}}{r_i}-v_i^2}\sim \frac{\mu_{3}}{\frac{2\mu_{3}}{r_i}-v_i^2}\label{eq:a}\\
{e_i}^2 &=& 1+\frac{2l_i^2\epsilon_i}{\mu_{3}^2}\label{eq:e} 
\end{eqnarray}
where $\mathbf{l}_i = \mathbf{r}_i\times\mathbf{v}_i$ and $\epsilon_i\sim -\frac{\mu_{3}}{2a_i}$ are the specific angular momentum and specific energy, respectively. \\

\subsection{Verification of the analytical results with few-body scattering experiments}\label{sec:verify}

To validate the correctness of Equations~\ref{eq:v1p}-\ref{eq:v2p}, we set up a suite of scattering experiments with the few-body code {\tt SpaceHub} \citep{2021Spacehub}, and compare the numerical results with the analytical solutions.   
We perform our tests by adopting numerical values for the masses to represent the astrophysical scenario of stellar mass BHs scattering in the potential of a supermassive BH (Case A, AGN scenario), and of planets in the potential of a central star (Case B, planetary disk). The corresponding numerical values are, respectively,
Case A: $m_1/m_3 = 10^{-8}$, $m_2/m_3 = 3\times10^{-7}$ and, and Case B:  $m_1/m_3 = m_2/m_3 =3\times10^{-6}$.

We verify the simplest retrograde circular scattering where both $m_1$ and $m_2$ orbit around $m_3$ in circular orbits in the opposite direction.  We also test the  results for scattering locations at different distances $r$ from the central object $m_3$ to verify that length in this problem can be scaled freely with Hill radius.

\begin{figure}
\includegraphics[width=\columnwidth]{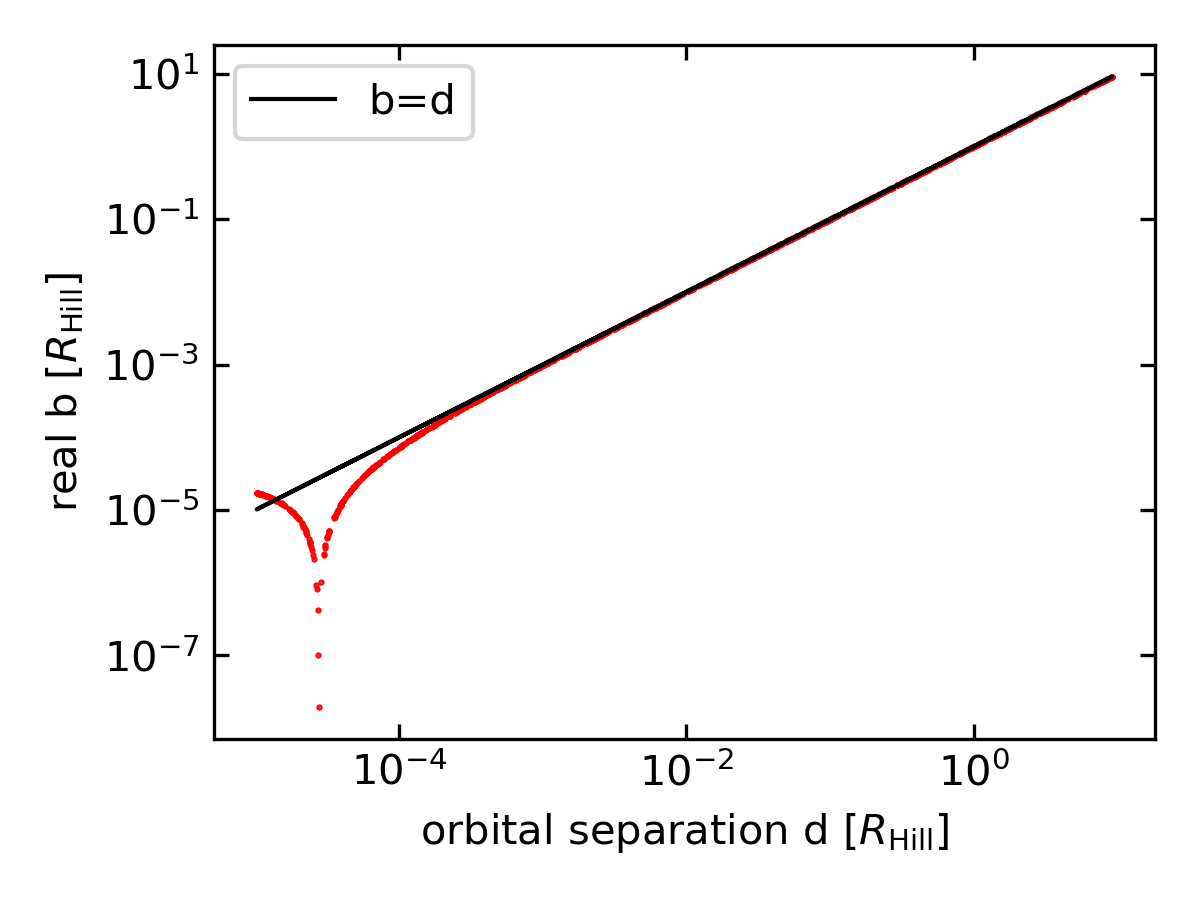}
\caption{The relationship between semi-major axis difference and real impact parameter for circular orbit scatterings. The masses are $m_1=30M_\odot$, $m_2=1M_\odot$ and $m_3=10^8M_\odot$. This relationship is scale-free with $r$.}
\label{fig:b-d relation}
\end{figure}

For retrograde circular scatterings, the initial phase difference between $m_1$ and $m_2$ is $\pi$ and the impact parameter (in this case, the semi-major axis difference) is $d$. For every single scattering, the simulation stops when the post-scattered relative phase between $m_1$ and $m_2$ becomes $\pi$ again. We found that the semi-major axis difference may not be a good approximation of the impact parameter when $d<10^{-4}R_{\rm Hill}$ as indicated by Figure~\ref{fig:b-d relation}. Therefore, we run a set of simulations to obtain the relationship between the semi-major axis difference and the real impact parameter. The real impact parameters are calculated from the closest approach and relative velocity between $m_1$ and $m_2$ obtained from the simulations.

We do not test prograde scatterings with zero eccentricity because from the simulations we found that prograde circular scatterings result in continuous scatterings in which $m_1$ and $m_2$ continuously swap their positions after the first close approach. This makes it difficult to distinguish individual scattering between these continuous scatterings. This limitation will be discussed in the last subsection.

\begin{figure}
\includegraphics[width=.95\columnwidth]{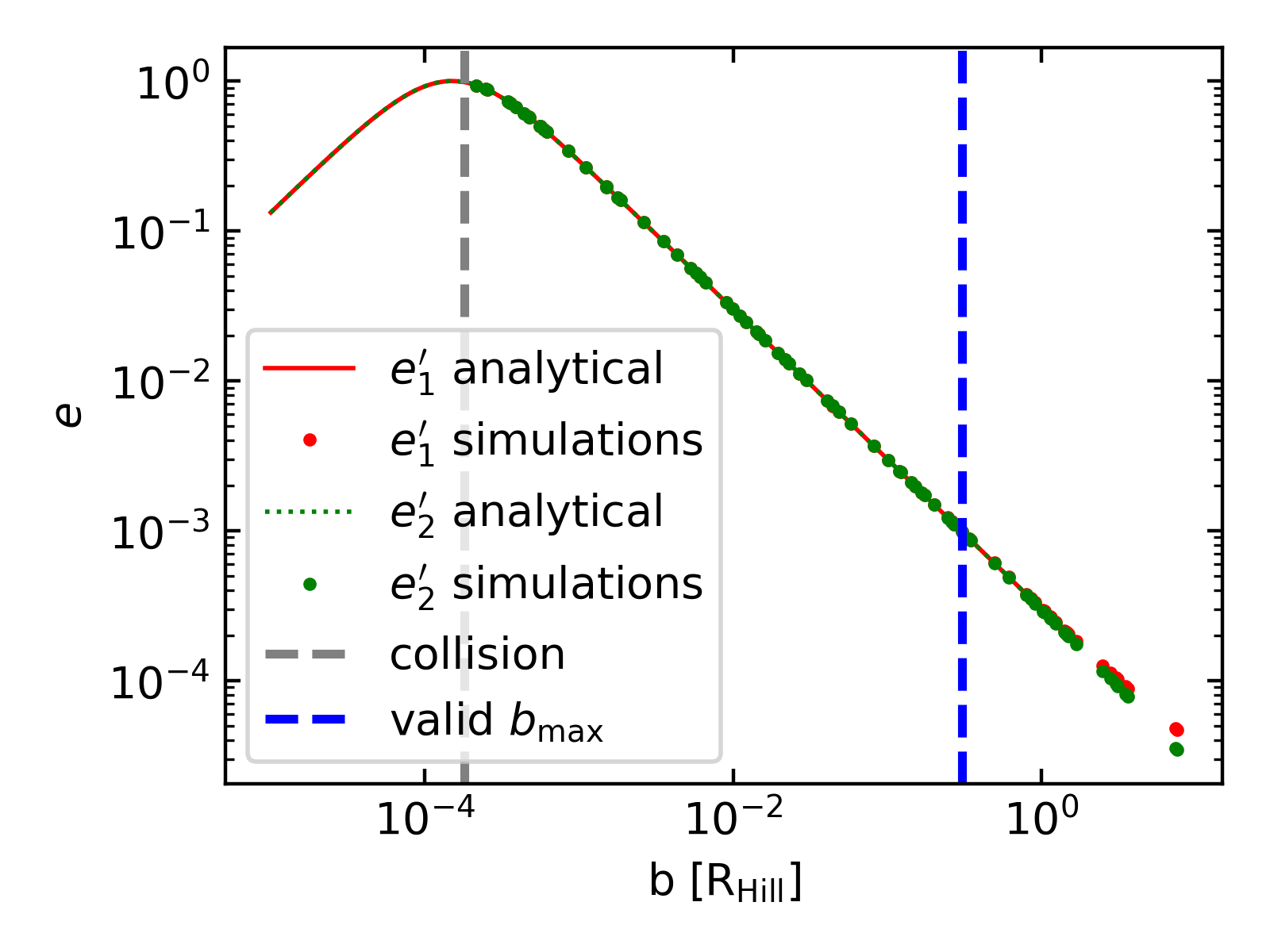}\\
\includegraphics[width=.95\columnwidth]{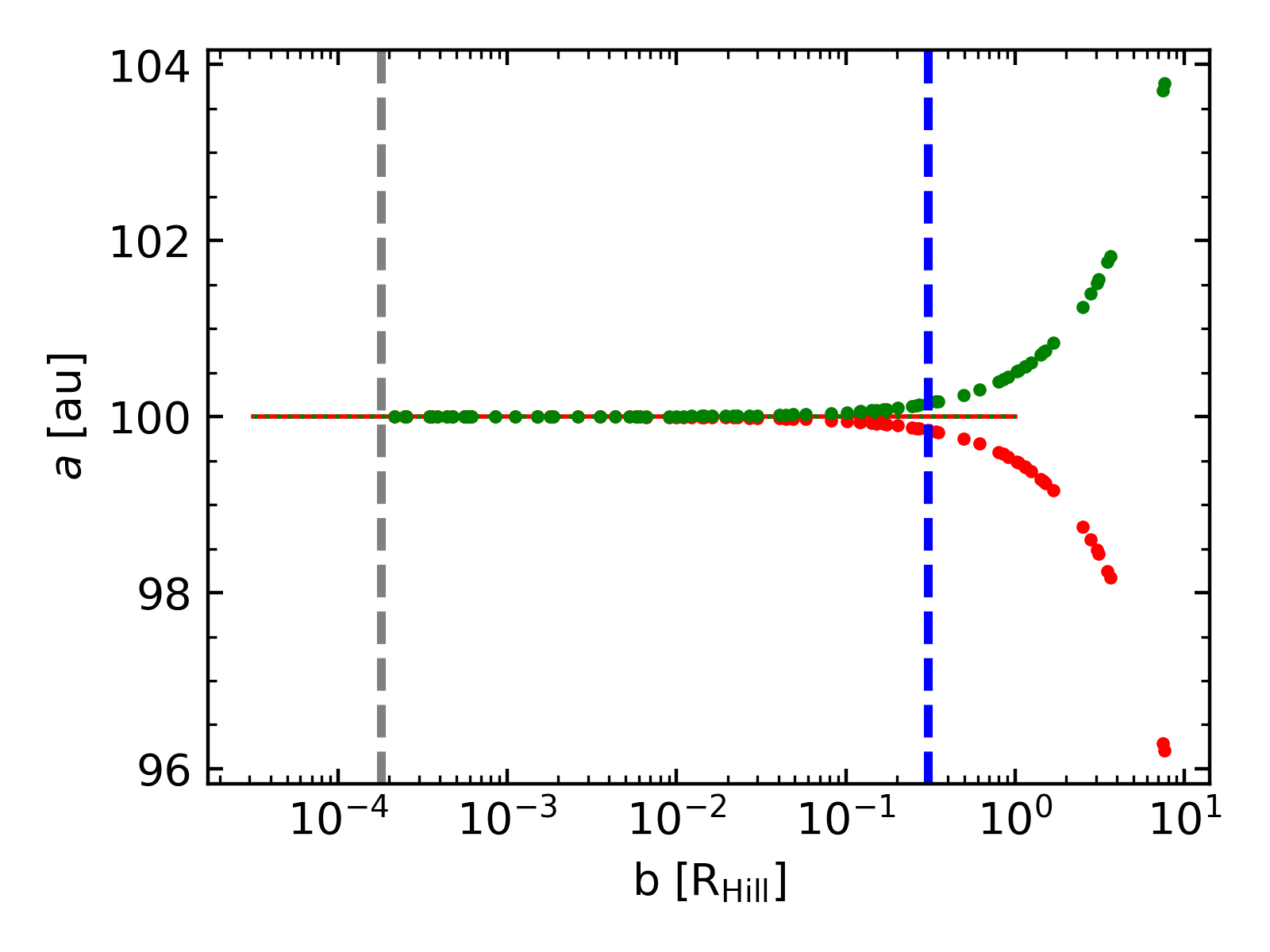}
\caption{Post-scattered semi-major axis and eccentricity of circular retrograde scatterings from simulations and analytical solution given by this paper. The scattering setup is  $m_1 = 1 M_\oplus$, $m_2=1M_\oplus$ and $m_3=1M_\odot$. The gray vertical lines indicate the impact parameter where $m_1$ and $m_2$ collide with each other (assuming earth-density planets). Vertical blue lines indicate the maximum impact parameter for which the analytical approximation is valid. }
\label{fig:verify-planet-ret}
\end{figure}

Figure~\ref{fig:verify-planet-ret} shows, for the study case B, the comparison between the analytical approximation given by Equation~\ref{eq:v1p}, \ref{eq:v2p}, \ref{eq:a}, \ref{eq:e} and the numerical simulations. The left panels show the post-scattered eccentricity of $m_1$ and $m_2$ while the right panels indicate the post-scattered semi-major axis. This figure indicates that for retrograde circular scatterings with this mass combination, our analytical approximation can perfectly describe the two-body scattering for a region of impact parameters ranging from the minimum collision value (indicated by the vertical gray line) to the maximum value. The simulation results indicate that even for bigger impact parameters $b>b_{\rm valid, max}$, the analytical approximation still describes the post-scattered eccentricity and semi-major axis very accurately.

\begin{figure}
\includegraphics[width=.95\columnwidth]{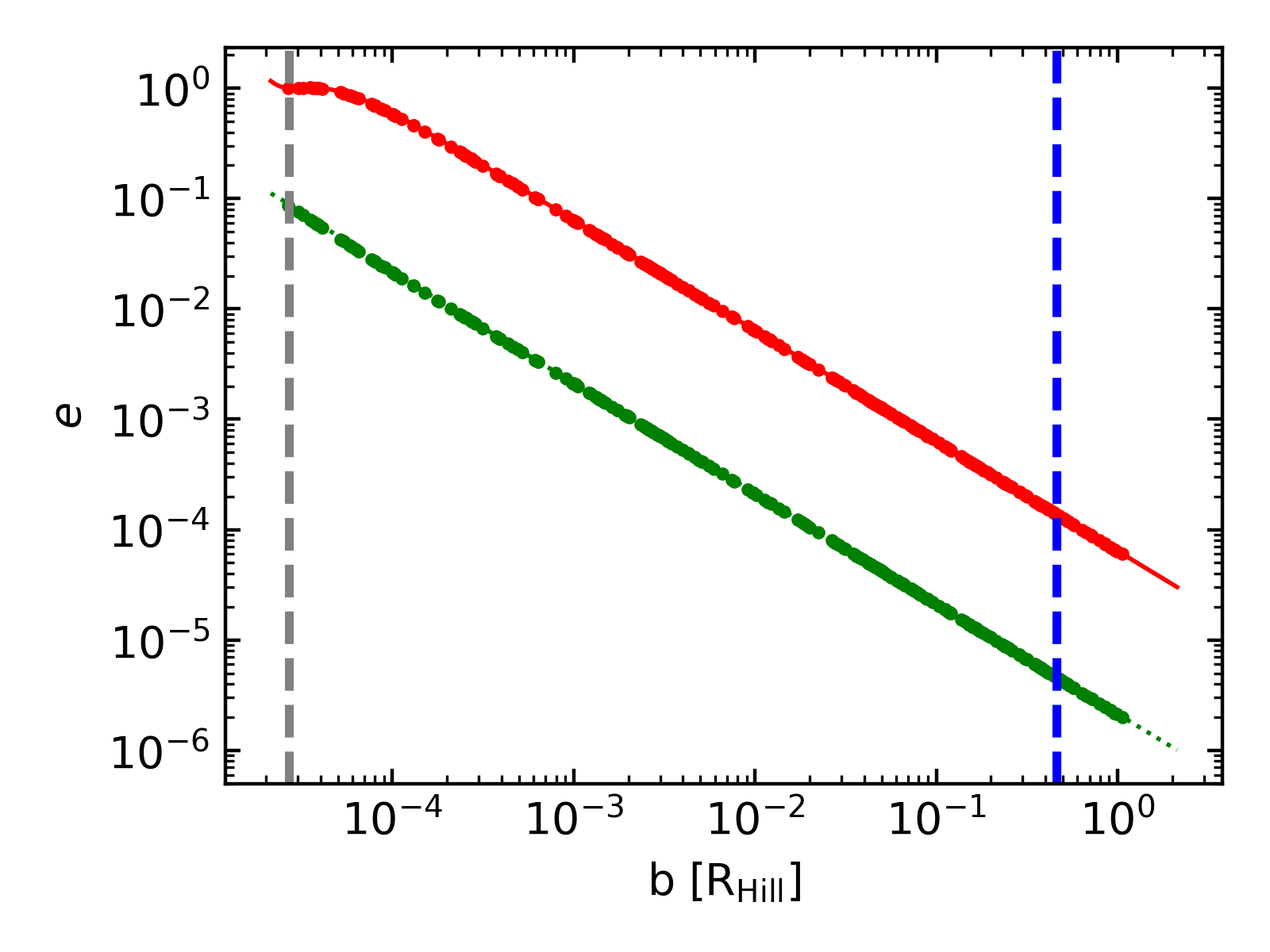}\\
\includegraphics[width=.95\columnwidth]{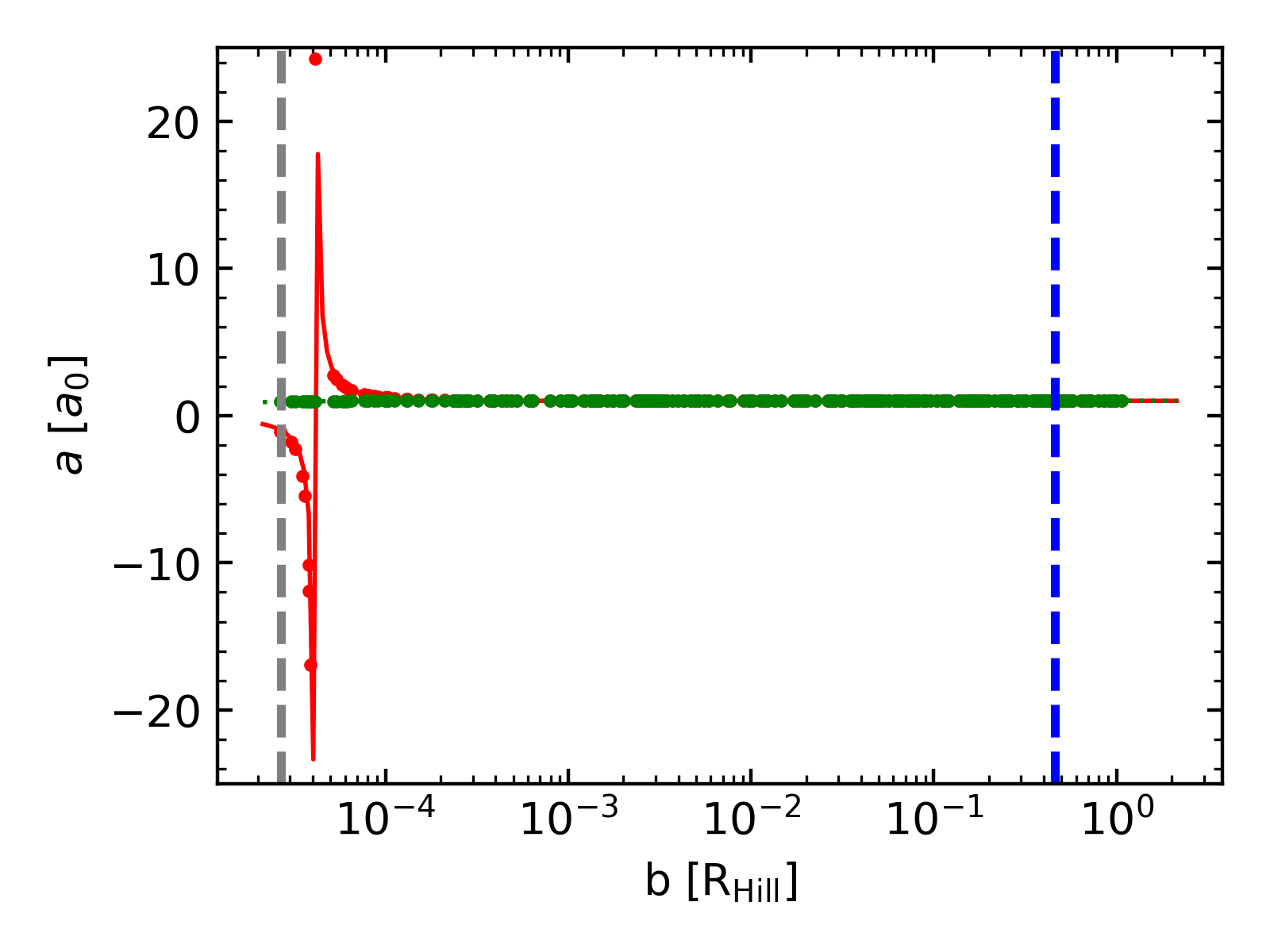}
\caption{Same as Figure~\ref{fig:verify-planet-ret} but for case A where $m_1 = 1 M_\odot$, $m_2=30M_\odot$ and $m_3=10^8M_\odot$. }
\label{fig:verify-smbh-ret}
\end{figure}

Figure~\ref{fig:verify-smbh-ret} shows the same comparison but for the larger mass set of case A. The post-scattered eccentricities and semi-major axis from the simulations fit the analytical results. 

\subsection{Impact parameter for a given closest approach}\label{sec:rmin}

For a given closest approach $R_{\rm min}$ between $m_1$ and $m_2$, the corresponding impact parameter for this closest approach is 
\begin{eqnarray}
b_{\rm min} &=& R_{\rm min}\sqrt{1+\frac{2Gm_{12}}{v_\infty^2 R_{\rm min}}}\,,
\end{eqnarray}
where $v_\infty$ is the pre-scattering Keplerian velocity difference determined by the  orbital parameters  of $m_1$ and $m_2$ at the scattering location $r$ with $a_1=a_2=r$ and $e_2=0$,
\begin{equation}
v_\infty^2=
\left\{
\begin{aligned}
    &{2\frac{Gm_3}{r}}\left(1-\sqrt{1-e_1^2}\right)\quad {\rm I\&II}\\
    &{2\frac{Gm_3}{r}}\left(1+\sqrt{1+e_1^2}\right)\quad {\rm III\&IV}.\\ 
\end{aligned}
\right.
\end{equation}
Then $b_{\rm min}$ can be expressed as
\begin{equation}
    b_{\rm min} = R_{\rm min}\sqrt{1+\frac{m_{12}}{m_3}\frac{r}{R_{\rm min}}\frac{1}{\Delta_\pm}}
\end{equation}
where $\Delta_\pm=1\pm\sqrt{1\pm e_1^2}$. 
This expression has the following limits:
\begin{equation}
    b_{\rm  min} =
    \left\{
    \begin{aligned}
    &\bigg(\frac{m_{12}}{m_3\Delta_\pm}\bigg)^{1/2}\bigg(\frac{r}{R_{\rm min}}\bigg)^{1/2}R_{\rm min}, \quad r\gg r_c\\
    &R_{\rm min}, \quad r\ll r_c
    \end{aligned}
    \right.
\end{equation}
where $r_c =R_{\rm min}\Delta_\pm\frac{m_3}{m_{12}}$. This is useful to obtain the corresponding impact parameter if a given closest approach is required (i.e. collision between $m_1$ and $m_2$, or $m_1$ is tidally disrupted by $m_2$).

\subsection{Ejection of the small object}\label{sec:ejection}
The scattering between $m_1$ and $m_2$ can eject either $m_1$ and $m_2$ from the potential of $m_3$. To calculate the critical impact b for ejecting $m_1$, we can plug the Keperian velocities into Equation~\ref{eq:v1p}-\ref{eq:v2p} and Equation~\ref{eq:a} and solve $e^\prime = 1$. Solving the general case with arbitrary $e_1$ is not easy. However, it's relatively easy to get the solution in  the limit of $e_1\rightarrow 0$,
\begin{eqnarray}\label{eq:ej-b}
b/r=\left\{
\begin{aligned}
&{\rm no\quad solution}, {\rm I\& II}\\
&\pm\frac{\sqrt{7m_2^2-10m_1m_2-m_1^2}}{4m_3},{\rm III\&IV}\,.
\end{aligned}
\right.
\end{eqnarray}
This requires $-m_1^2-10m_1m_2+7m_2^2>0$ for a physical solution, that is ${m_2}/{m_1}>\frac{5+4\sqrt{2}}{7}$.
Similar results can be obtained in the limit of $e_1\rightarrow1$,
\begin{eqnarray}\label{eq:ej-b2}
b/r=\left\{
\begin{aligned}
&\frac{2m_2\pm\sqrt{(m_2-m_1)(7m_2+m_1)}}{2m_3},{\rm I\&IV}\\
&\frac{-2m_2\pm\sqrt{(m_2-m_1)(7m_2+m_1)}}{2m_3}, {\rm II\&III}
\end{aligned}
\right.
\end{eqnarray}
The above conditions require the mass ratio $m_2/m_1$ to be
\begin{eqnarray}
\left\{
\begin{aligned}
&1<\frac{m_2}{m_1},{\rm I\&IV}\\
&\frac{3+2\sqrt{3}}{3}<\frac{m_2}{m_1},{\rm II\&III}\,.
\end{aligned}
\right.
\end{eqnarray}

\subsection{Direct collision with the massive object}\label{sec:tde}
The scattering between $m_1$ and $m_2$ can also lead to a direct collision with $m_3$. This case requires the post-encounter angular momentum of the small object to be zero.

\begin{figure}
\includegraphics[width=\columnwidth]{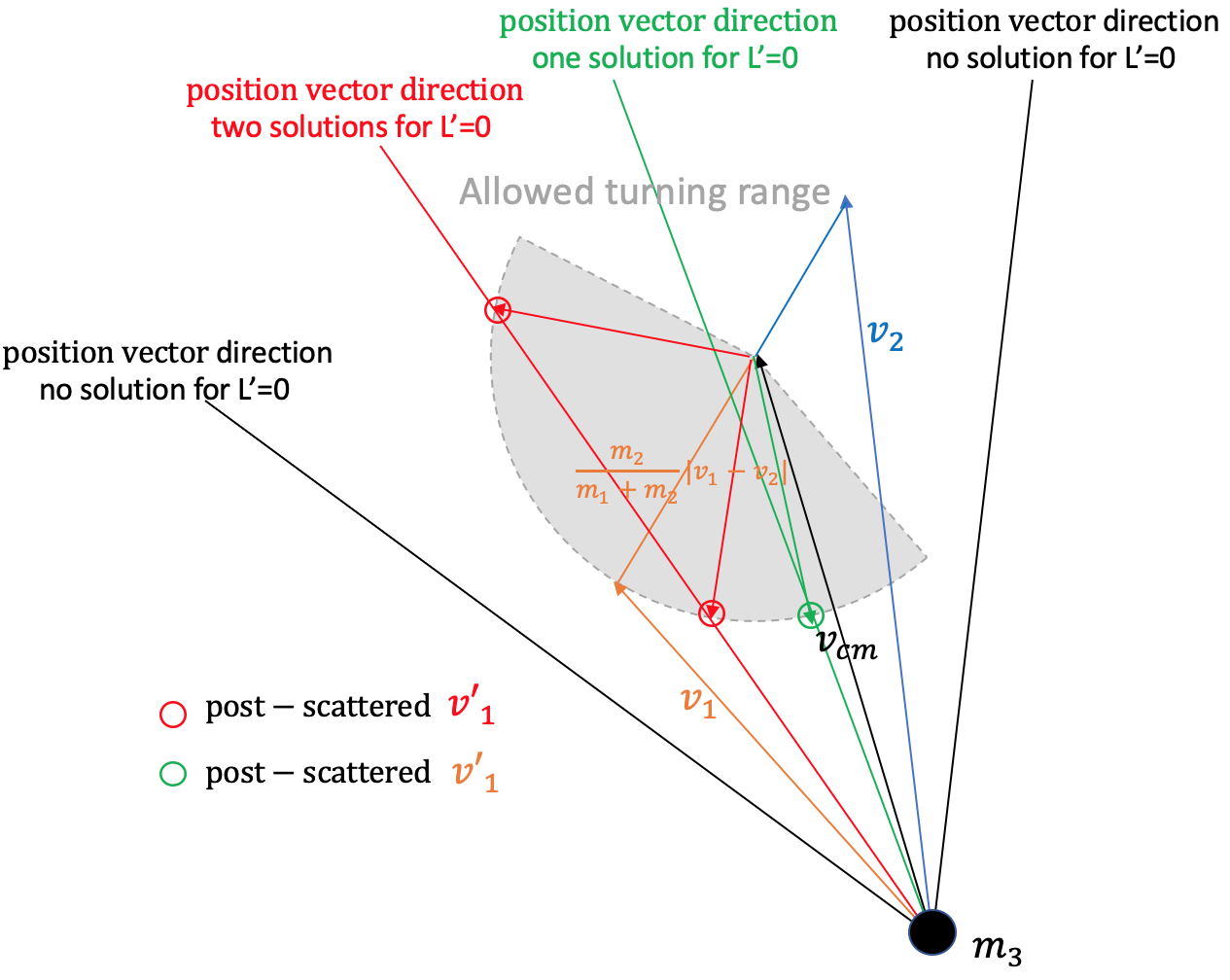}
\caption{Illustration of scatterings that make the post-scattering angular momentum of $m_1$, $L^\prime=r^\prime\times v^\prime$, equal to zero. This would lead to $m_1$ colliding with $m_3$. $v_1$ and $v_2$ are the initial velocities of $m_1$ and $m_2$.   }
\label{fig:l=0}
\end{figure}
Figure~\ref{fig:l=0} illustrates the possible solution for the post-encounter angular momentum of small objects to be zero. If the initial position vector (unchanged during the scattering) lays outside of the allowed turning region, it is impossible for the post-encounter velocity vector to be aligned with the position vector. Thus, no solution can be obtained for $L^\prime=0$. If there are two intersections between the position vector and the allowed turning region, there will be two solutions of b with $L^\prime=0$. Of course, if only one intersection can be found, then there is only one solution for b with zero post-encounter angular momentum.

The solutions for the four types of scattering are
\begin{eqnarray}
    b/r=\left\{
    \begin{aligned}
    &{\rm no\quad physical\quad solution},{\rm I}\\
    &\frac{m_{12}}{2m_3},\frac{m_2}{m_3}\frac{1}{\sqrt{1-e_1^2}},{\rm II:}(\sqrt{1-e_1^2}\ll1)\\
    &\frac{\sqrt{(3m_2-m_1)(m_1+m_2)}}{4m_3},{\rm III}\\
    &\frac{m_2}{m_3}\frac{1}{\sqrt{1-e_1^2}},{\rm IV:}(\sqrt{1-e_1^2}\ll1)\\
    &\frac{\sqrt{(3m_2-m_1)(m_1+m_2)}}{4m_3},{\rm IV:}(e_1\ll1)\\
    \end{aligned}
    \right.
\end{eqnarray}
 Type I scatterings correspond to the case indicated by the black line (no solution). Type III and IV scatterings are indicated by the green line (one solution), while type II scatterings are represented by the red line (two solutions).

\subsection{Examples of post-encounter orbital properties}
In this section we show the post-encounter orbital properties for the two cases we introduced in Sec~\ref{sec:verify}. These situations will be discussed further as astrophysical scenarios in Sec.~3.

\begin{figure*}
\includegraphics[width=2\columnwidth]{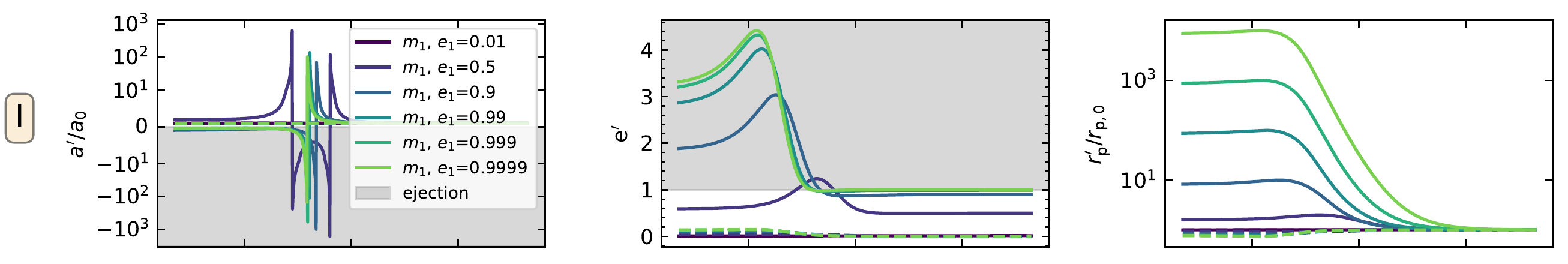}\\
\includegraphics[width=2\columnwidth]{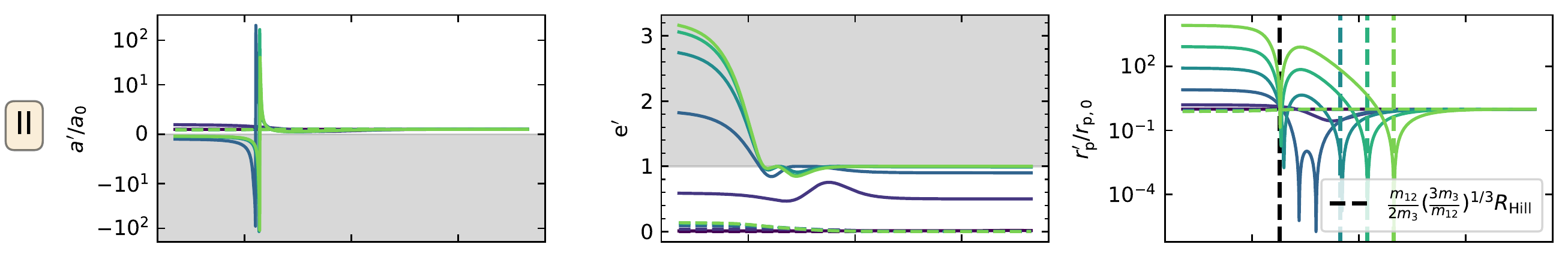}\\
\includegraphics[width=2\columnwidth]{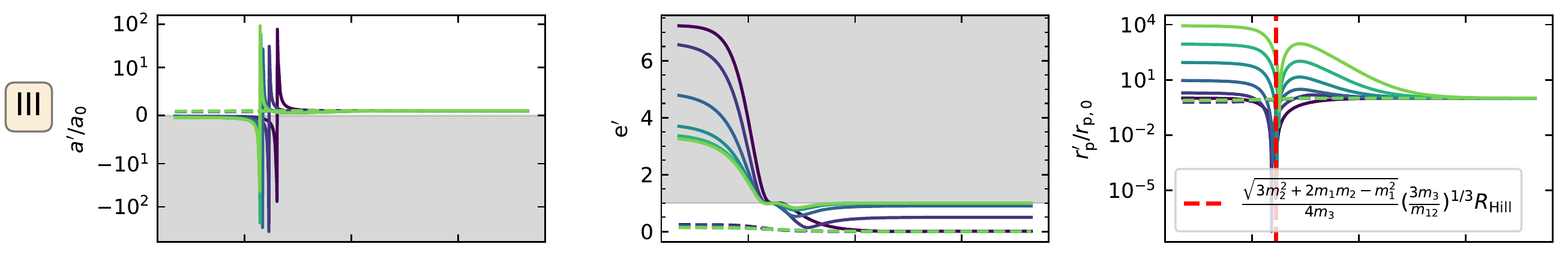}\\
\includegraphics[width=2\columnwidth]{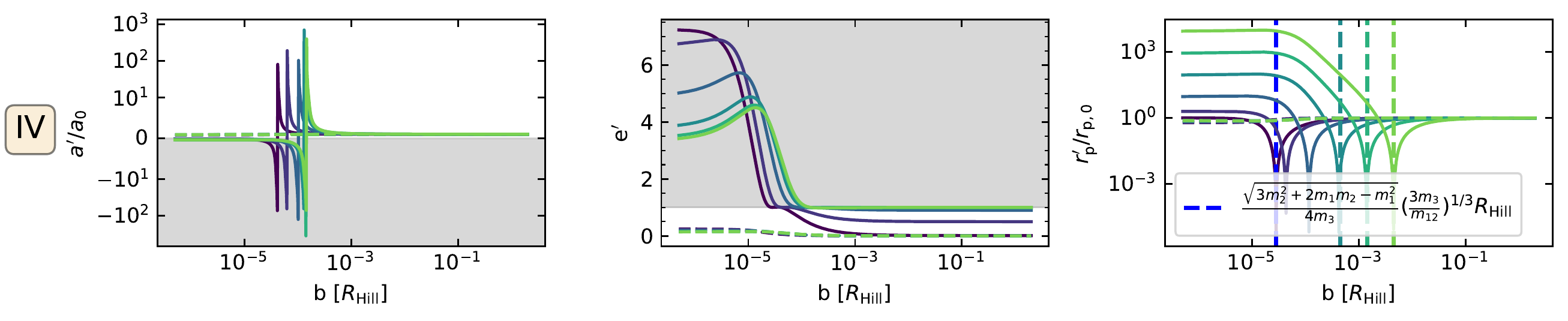}
\caption{Post scattered orbital properties of $m_1$ (solid lines) and $m_2$ (dashed lines) from scatterings in the case with $m_1 = 1 M_\odot$, $m_2=30M_\odot$ and $m_3=10^8M_\odot$. Note that these curves are independent of the scattering position $r$ to $m_3$. Color dashed vertical lines that are not labeled are $\frac{m_2}{m_3}\frac{1}{\sqrt{1-e_1^2}}\bigg(\frac{m_3}{m_{12}}\bigg)^{1/3}R_{\rm Hill}$.}
\label{fig:SMBH}
\end{figure*}

Figure~\ref{fig:SMBH} shows the post-scattered semi-major axis and eccentricity of the lighter object $m_1$ in case A for the different types of scatterings I, II, III, and IV shown in Figure~\ref{fig:schematics}. This case is a good example of scattering between a  main sequence star and a stellar-mass black hole around a supermassive black hole. It indicates that  type I and IV generally increase the semi-major axis of the lighter object $m_1$ until the impact parameter is down to $10^{-4}R_{\rm Hill}$ where ejection of $m_1$ starts to appear. In case I, $m_1$ transfers energy to $m_2$ but gains angular momentum from $m_2$. In case IV, $m_1$ transfers both energy and angular momentum to $m_2$. For type II scatterings, the semi-major axis of $m_1$ typically shrinks by a maximum factor of two if $m_1$ is in an extremely eccentric orbit. In this case, $m_1$ obtains energy from $m_2$ until  ejection. For type III scatterings, low eccentricity orbits tend to increase the semi-major axis while high eccentricity orbits tend to decrease the semi-major axis.

\begin{figure*}
\includegraphics[width=2\columnwidth]{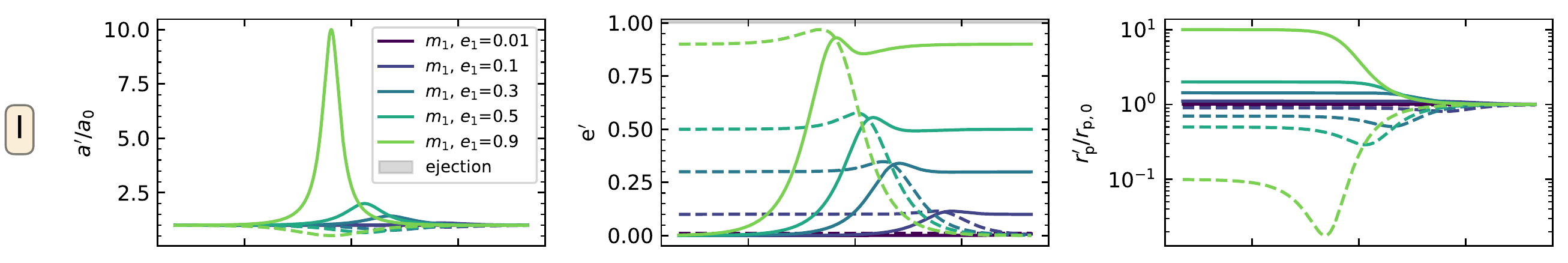}\\
\includegraphics[width=2\columnwidth]{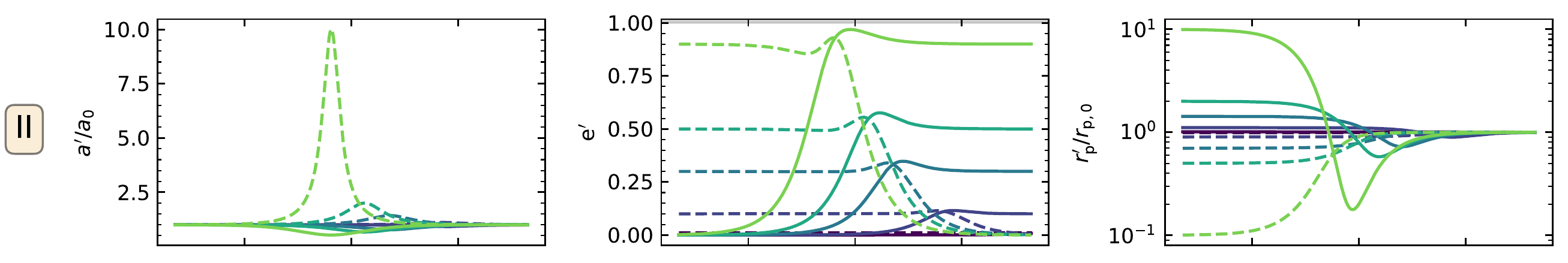}\\
\includegraphics[width=2\columnwidth]{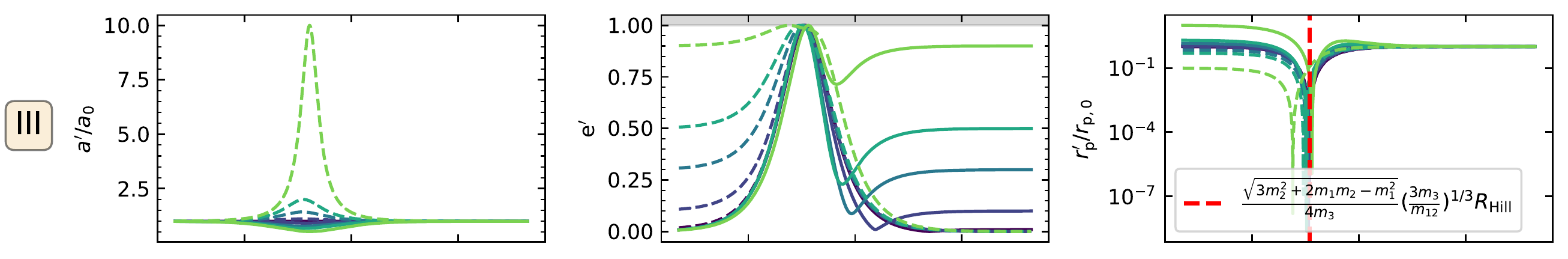}\\
\includegraphics[width=2\columnwidth]{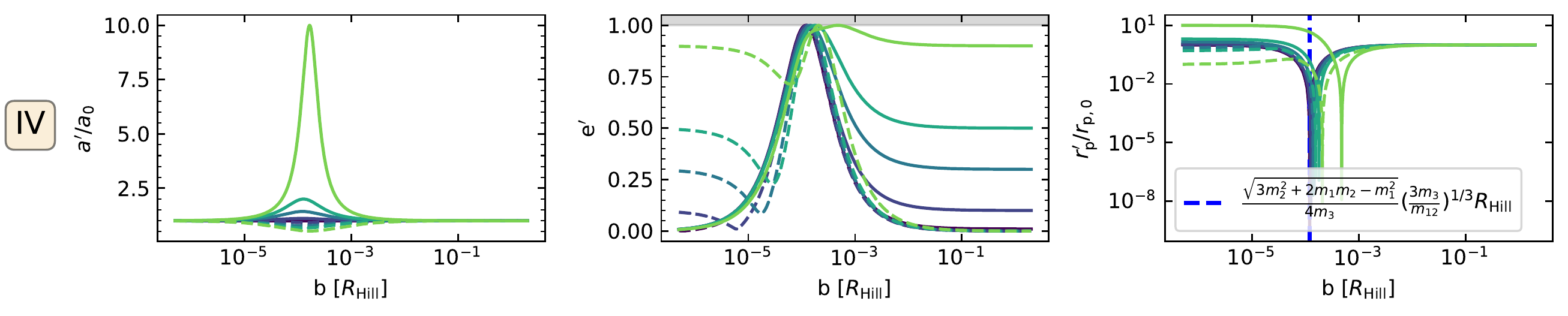}
\caption{Similar to Figure~\ref{fig:SMBH}. Post scattered orbital properties of $m_1$ and $m_2$ from scatterings  in the case with $m_1 = 1 M_\oplus$, $m_2=1M_\oplus$ and $m_3=1M_\odot$. The initial eccentricity of $m_2$ is always 0.}
\label{fig:planet-ae}
\end{figure*}
For equal mass scatterings (case B), as shown in Figure~\ref{fig:planet-ae}, type I\& II are symmetric between $m_1$ and $m_2$ in semi-major axis change, and type III\&IV show similar behaviors. For type I and II scatterings, maximum semi-major axis and eccentricity are achieved at the same impact parameter $b$ while in type III and type IV the maximum semi-major axis and maximum eccentricity occur at different impact parameters. Interestingly, for type I and II scatterings, low initial eccentricity cases can achieve maximum semi-major axis change with much larger impact parameters while in type III and IV scatterings, maximum semi-major axis change presents around $10^{-4}R_{\rm Hill}$ for all initial eccentricities. Because the mass ratio between $m_1$ and $m_2$ is unity, based on the calculation in Section~\ref{sec:ejection}, no ejection can be obtained in the  case.

\section{Applications}\label{sec:app}

In the following, we will discuss the direct astrophysical implications of our analytical results for the two cases mentioned above: {\em (i)} Stellar-mass BH-star scatterings under the potential of a central SMBH (such as, for example, the disk of an AGN); {\em (ii)} planet-planet scattering in the potential of a host star, which is the typical situation of a planetary system with coplanar planetary orbits.

\subsection{Micro-TDE in the presence of a central potential }

If, during the scattering of a stellar-mass BH of mass $m_2$ and a main sequence star of mass $m_1$ and radius $R_*$, the star gets within a distance
$r_{\rm t}=R_*(\frac{m_1}{3m_3})^{1/3}$ of the BH, the tidal force from the BH will tidally disrupt the star, giving rise to long X-ray/gamma-ray flares (e.g. \citealt{Perets2016}).

Plugin $r_{t}$ as $R_{\rm min}$ into Section~\ref{sec:rmin}, we can directly obtain the critical impact parameter for a micro-TDE
\begin{eqnarray}
b_{\rm \mu TDE} &=& R_{\rm \mu TDE}\sqrt{1+\frac{2G(m_{12})}{v_\infty^2 R_{\rm \mu TDE}}}\,.
\end{eqnarray}
The cross-section of this micro-TDE can thus be estimated via 
\begin{eqnarray}
    \sigma_{\rm \mu TDE} &=& \int_0^{b_{\rm \mu TDE}}db  = b_{\rm\mu TDE} \\
    &=& \bigg(\frac{m_2}{3m_1} \bigg)^{1/3}R_*\sqrt{1+ \bigg(\frac{m_{12}}{m_3}\bigg)\bigg(\frac{3m_1}{m_2}\bigg)^{1/3}\frac{r}{R_*}\frac{1}{\Delta_\pm}}.
\end{eqnarray}
This expression has the following limits:
\begin{equation}
    \sigma_{\rm \mu TDE} =
    \left\{
    \begin{aligned}
    &\bigg(\frac{m_{12}}{m_3\Delta_\pm}\bigg)^{1/2}\bigg(\frac{r}{R_{\rm \mu TDE}}\bigg)^{1/2}R_{\rm \mu TDE}, \quad r\gg r_c\\
    &R_{\rm \mu TDE}, \quad r\ll r_c
    \end{aligned}
    \right.
\end{equation}
where $r_c =R_*\Delta_\pm\bigg(\frac{m_3}{m_{12}}\bigg)\bigg(\frac{m_2}{3m_1}\bigg)^{1/3} $. Note that we have constructed a cross-section with units of length, rather than area, since our scattering problems are co-planar, and hence
2-dimensional.

\begin{figure*}
\includegraphics[width=.95\columnwidth]{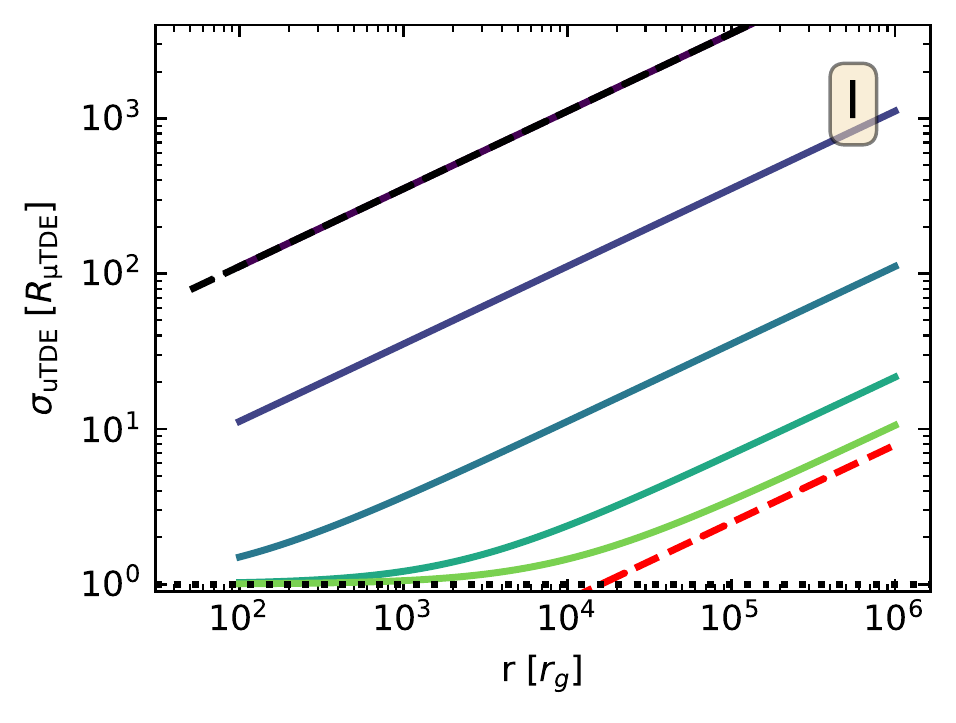}
\includegraphics[width=.95\columnwidth]{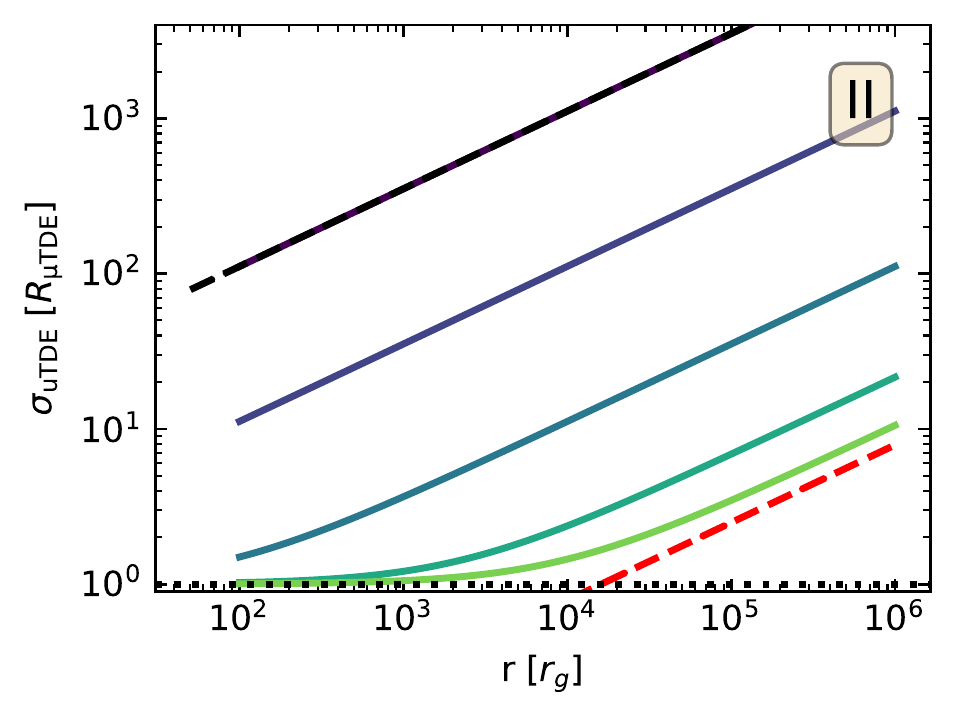}\\
\includegraphics[width=.95\columnwidth]{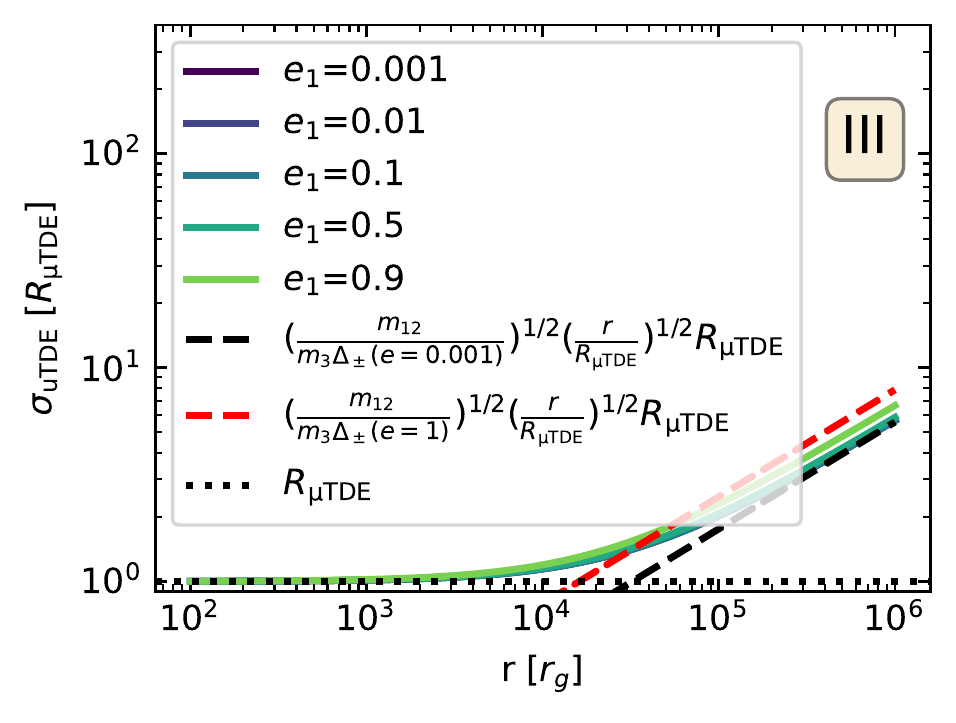}
\includegraphics[width=.95\columnwidth]{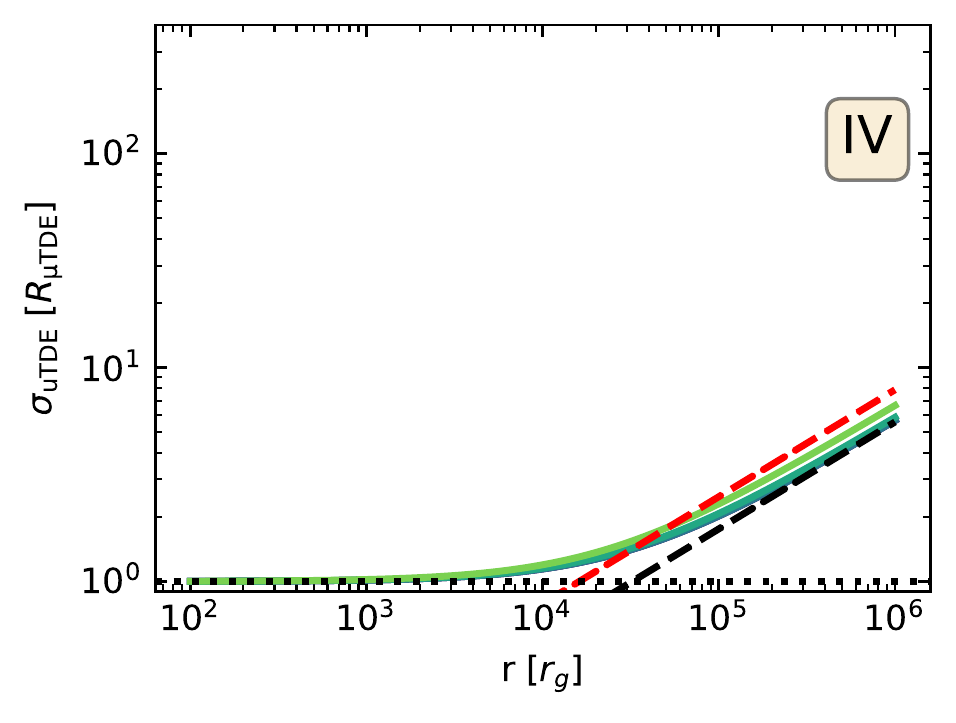}
\caption{Micro-TDE (disrupted by the stellar mass BH) cross-section as a function of $r$ for different types of scatterings.  The mass of the central SMBH is $10^8$  $M_\odot$ and the mass of the star is $M_\odot$ and the mass of the stellar mass BH that tidally disrupts the star is $30M_\odot$.}
\label{fig:utde}
\end{figure*}

Figure~\ref{fig:utde} shows the cross-section of the micro-TDE as a function of $r$ for the four different types of scatterings shown in Figure~\ref{fig:schematics},  and for different orbital eccentricities of the star. Generally, the rate of micro-TDEs stays nearly constant if $r<r_c$ and increases as $r^{1/2}$ in the region of $r>r_c$. In the outer region, the orbital velocity difference between $m_1$ and $m_2$ around the SMBH $m_3$ is significantly smaller than in the inner region,  for both prograde scatterings (I and II) and retrograde scatterings (III and IV). Thus, the gravitational focusing effect in the larger $r$ region is stronger. Since the star will be destroyed at the fixed radius $R_{\rm \mu TDE}$, a stronger focusing effect in the larger $r$ region leads to a larger cross-section of the micro-TDE. This focusing effect is extremely strong in prograde circular scatterings as shown in the upper two panels of Figure~\ref{fig:utde}, which contributes most of the micro-TDEs in AGN disks.

\subsection{Star ejection during a BH-star scattering}

\begin{figure*}
\includegraphics[width=.95\columnwidth]{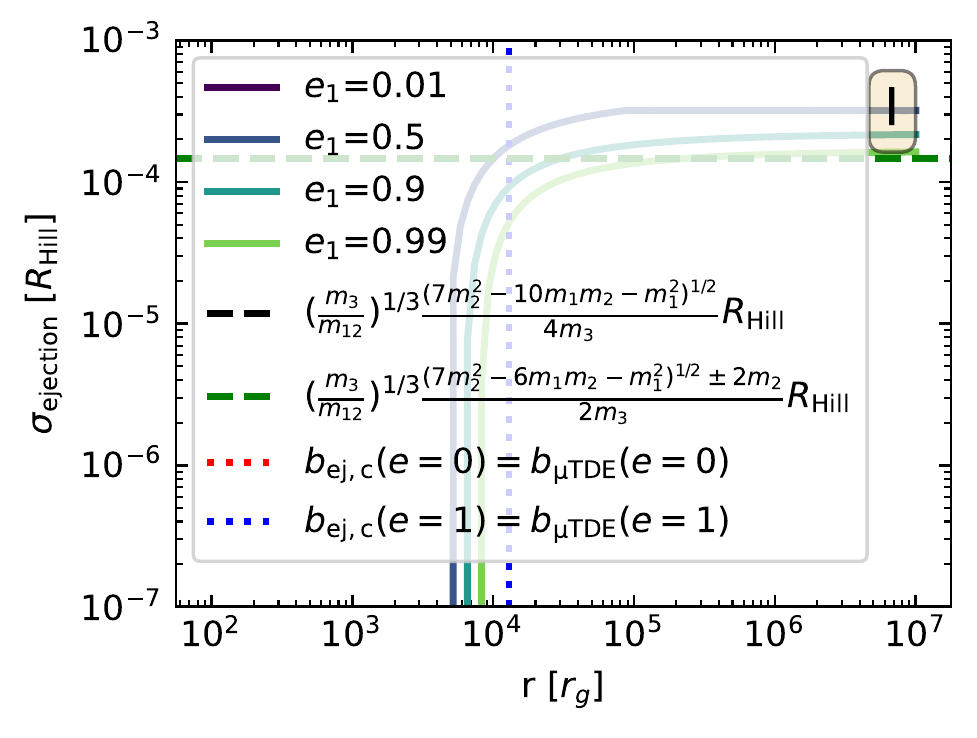}
\includegraphics[width=.95\columnwidth]{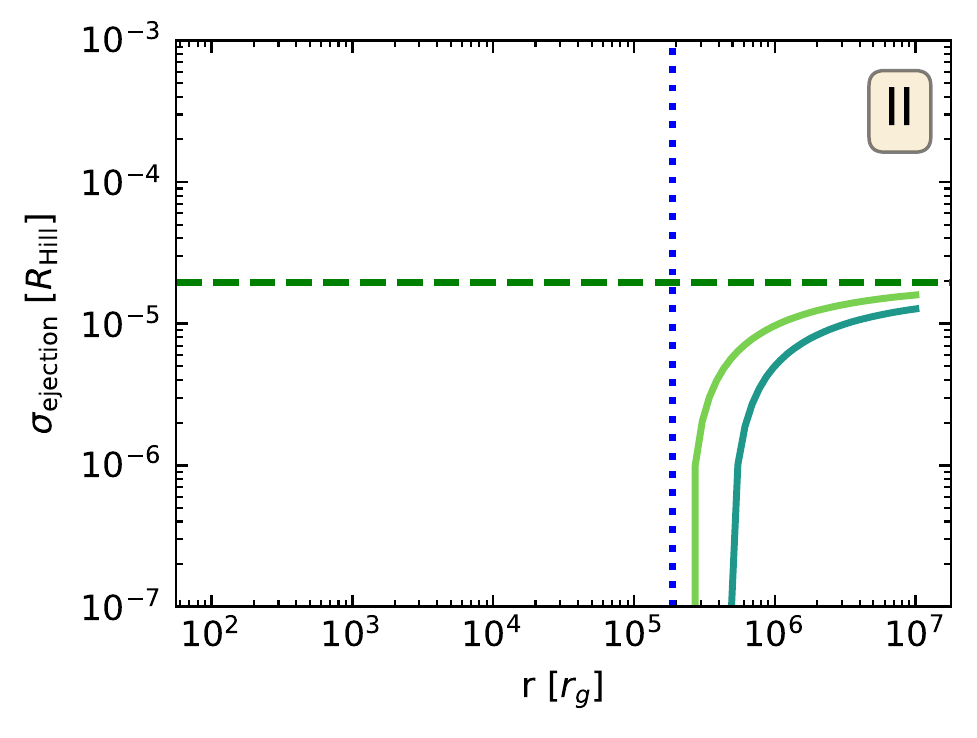}\\
\includegraphics[width=.95\columnwidth]{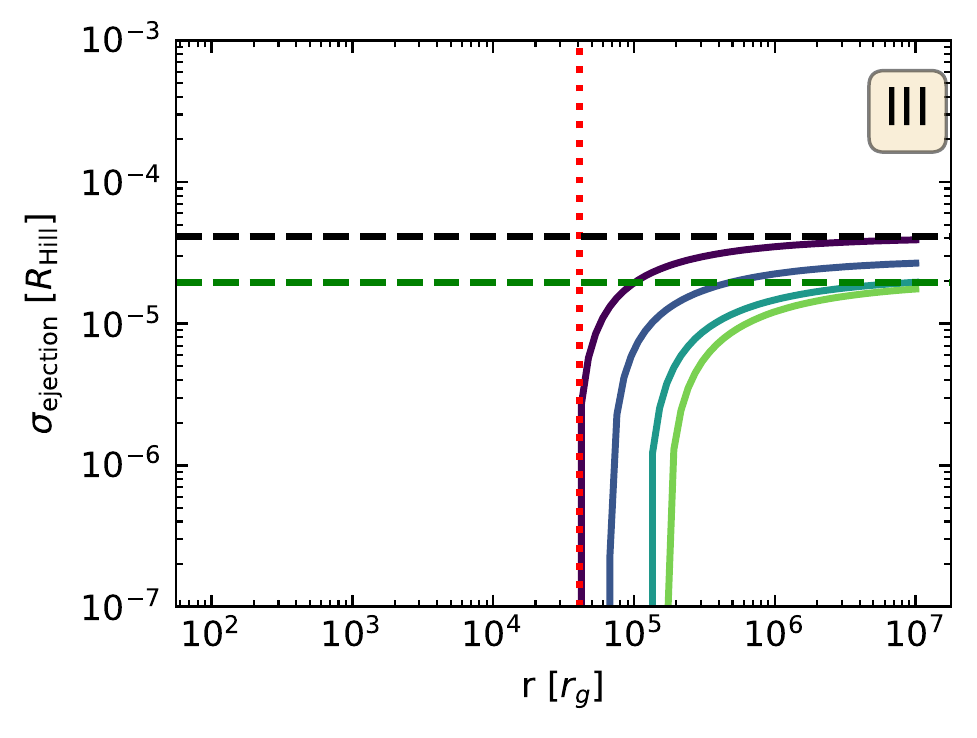}
\includegraphics[width=.95\columnwidth]{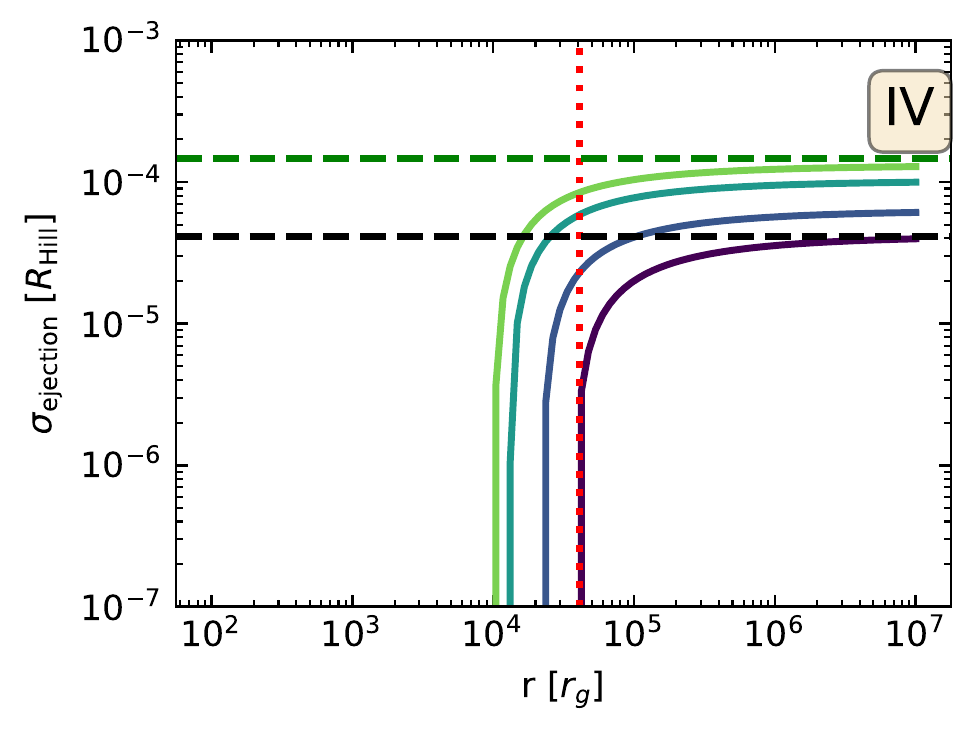}
\caption{Similar to Figure~\ref{fig:utde}, but for the star ejection. The horizontal dashed lines show the approximation from Equation~\ref{eq:ej-cross}. The vertical dotted lines indicate the critical $r$, where the star gets tidally disrupted by the stellar mass black hole before it gets ejected.}
\label{fig:ejecction}
\end{figure*}

Instead of being disrupted by the stellar mass BH or the SMBH, the star could also be unbound from the SMBH by the BH during the scattering. To eject the star, the post-scattered semi-major axis as calculated from Equation~\ref{eq:a} needs to be smaller than zero. Therefore, the cross-section of the star ejection is
\begin{equation}
    \sigma_{\rm ej} = \int_{\Sigma L_{\rm ej}} db\,,
\end{equation}
where $L_{\rm ej}$ encompasses all the regions in the parameter space of the impact parameter $b$ that give a post-scattered eccentricity of the star smaller than zero (i.e. the post-scattered orbit is hyperbolic).

From Section~\ref{sec:ejection}, we then obtain that the cross-section of star ejection in the limit of $e_1\rightarrow 0$ is
\begin{eqnarray}
\label{eq:ej-be}
&&\sigma_{\rm ej}({m_2}/{m_1}>\frac{5+4\sqrt{2}}{7})=\\
&&\left\{
\begin{aligned}
&0, {\rm I\& II}\\
&\bigg(\frac{3m_3}{m_{12}}\bigg)^{1/3}\frac{\sqrt{7m_2^2-10m_1m_2-m_1^2}}{4m_3}R_{\rm Hill}, {\rm III\&IV}
\end{aligned}
\right.\\
&&\sigma_{\rm ej}(1<{m_2}/{m_1}<\frac{5+4\sqrt{2}}{7})=0\,.
\end{eqnarray}

In the opposite limit of $e_1\rightarrow1$, the cross-section is
\begin{eqnarray}
&&\sigma_{\rm ej}(\frac{3+2\sqrt{3}}{3}<m_2/m_1)=\nonumber\\
&&\left\{
\begin{aligned}
&\frac{2m_2+\sqrt{(m_2-m_1)(7m_2+m_1)}}{2m_3}\bigg(\frac{3m_3}{m_{12}}\bigg)^{1/3}R_{\rm Hill},{\rm I\&IV}\\
&\frac{-2m_2+\sqrt{(m_2-m_1)(7m_2+m_1)}}{2m_3}\bigg(\frac{3m_3}{m_{12}}\bigg)^{1/3}R_{\rm Hill},{\rm II\&III}
\end{aligned}
\right.\label{eq:ej-cross}\\
&&\sigma_{\rm ej}(1<m_2/m_1<\frac{3+2\sqrt{3}}{3})=\nonumber\\
&&\left\{
\begin{aligned}
&\frac{\sqrt{(m_2-m_1)(7m_2+m_1)}}{m_3}\bigg(\frac{3m_3}{m_{12}}\bigg)^{1/3}R_{\rm Hill}, {\rm I\&IV}\\
&0, {\rm II\&III}.
\end{aligned}
\right.\label{eq:ej_cross2}
\end{eqnarray}

Figure~\ref{fig:ejecction} shows the ejection cross-section of the star for different types of scatterings and different initial orbital eccentricities of the star. For prograde scatterings in type I and II, due to the low relative velocity between $m_1$ and $m_2$ in nearly circular orbits, the gravitational focusing between $m_1$ and $m_2$ can be very strong, and hence stars with low eccentricity orbits can be easily disrupted by the BH before they acquire enough energy to be ejected. Only stars in a highly eccentric orbit, requiring less energy for ejection, can hence be ejected before the disruption. This is consistent with what we obtained from Equation~\ref{eq:ej-cross}.

For retrograde scatterings in type III and IV, due to the large relative velocity between $m_1$ and $m_2$, gravitational focusing is much weaker than for prograde scatterings.  Thus an encounter with a much smaller impact parameter can be achieved without star disruption. Therefore, even stars with low eccentricity orbits can be ejected by the stellar mass BH. The horizontal dashed lines show the approximation obtained from Equation~\ref{eq:ej-cross}, indicating that the ejection cross-section is effectively independent of $r$. The sharp vertical cutoffs mark where micro-TDEs take over. In the left small $r$ region, the required impact parameter for star ejection is smaller than the micro-TDE impact parameter. Therefore, the star will be tidally disrupted by the stellar mass BH and hence there is no ejection.

\subsection{Star tidally disrupted by the central SMBH (TDE)}

\begin{figure*}
\includegraphics[width=.95\columnwidth]{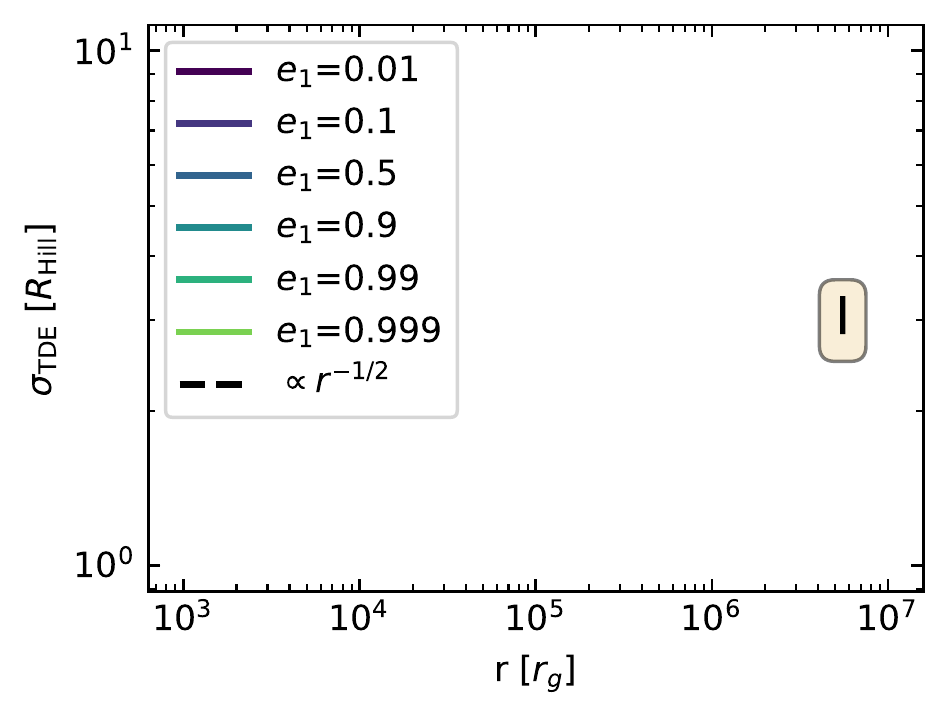}
\includegraphics[width=.95\columnwidth]{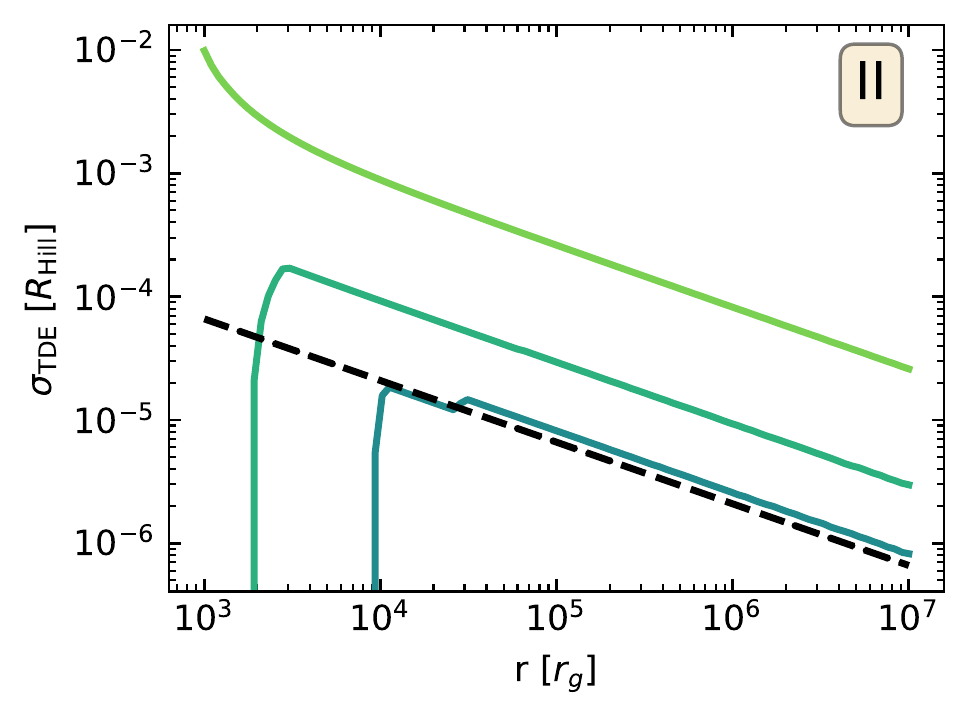}\\
\includegraphics[width=.95\columnwidth]{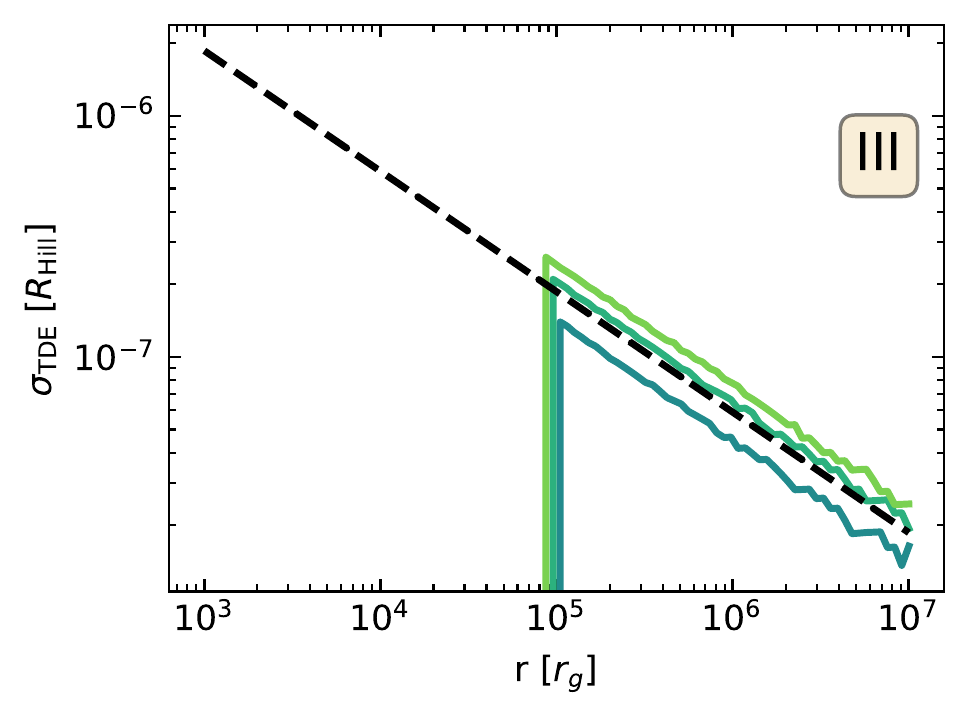}
\includegraphics[width=.95\columnwidth]{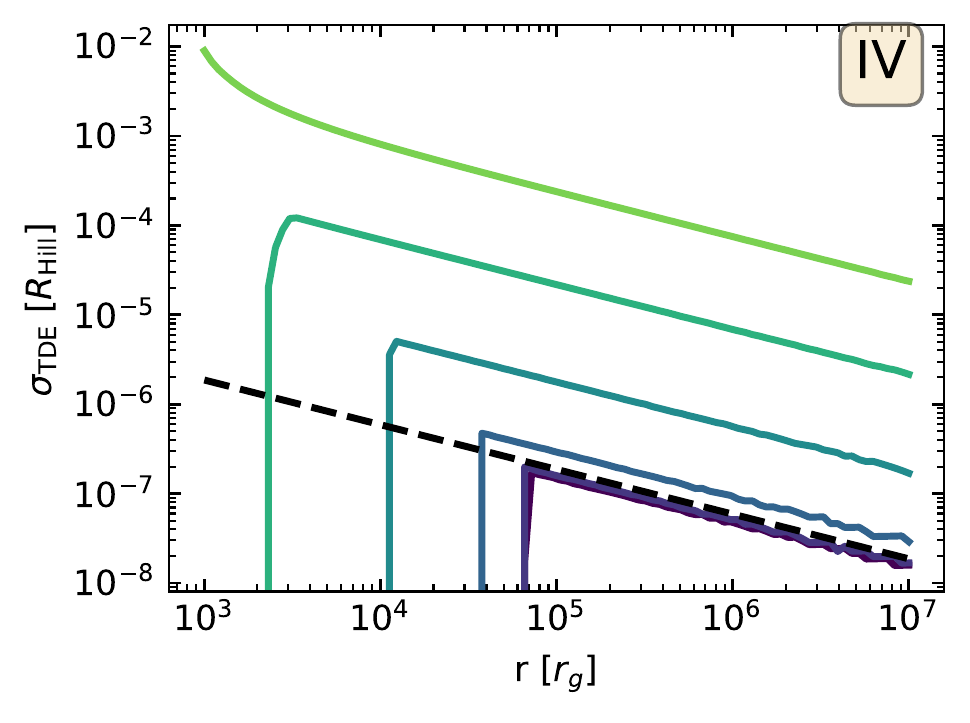}
\caption{TDE (disrupted by the SMBH) cross-section as a function of $r$ for different types of scatterings as shown in Figure~\ref{fig:schematics}}
\label{fig:tde}
\end{figure*}

`Standard' TDEs, in which the star is disrupted by the SMBH in quiescent galactic nuclei, have been extensively studied in the literature. The resulting flares are widely used to study the properties of the SMBH (mass, spin) as well as the populations of the host galactic nucleus \citep[e.g.][]{Bloom11}. TDEs in AGN disks are due to: either a) a 'standard' TDE in a nucleus where the TDE orbit crosses an AGN disk \citep{Cutoff17,Chan19} or b) due to a dynamical interaction between 2 or 3 bodies in the AGN disk that scatters the star onto the SMBH \citep{McKernan2022}\footnote{This picture is complicated by the very different evolution of stars embedded in AGN disks \citep{Cantiello2021, Jermyn2021} and which could have a significant impact on dynamical encounters within the disk and the disk itself \citep{Jermyn2022}.}. In case a) the TDE is due to the standard 2-body scattering into the loss cone independent of the disk and the rate of occurrence is the same as regular TDEs for that galaxy type. In case b) the TDE is due to 2-body or 3-body scattering in the AGN disk and the rate of occurrence of such events is a function of  disk size and the number of embedded objects within it. TDEs in AGN disks can create unique signatures because of the presence of the disk and must correspond to a source of AGN variability \citep{Graham17}. Note that in the discussion below we assume that scattered stars in AGN disks are $\sim 1M_{\odot}$. Since stars on prograde orbits within AGN disks can very rapidly grow to $O(100M_{\odot})$ \citep{Cantiello2021}, this corresponds to assuming the scattered star lies on an embedded retrograde orbit.

In our scattering model, the scatterings between $m_1$ and $m_2$ can result in a star orbit with a very small pericenter. The SMBH could tidally disrupt the star if this enters the tidal disruption radius of the SMBH,
\begin{equation}
    R_{\rm TDE}\sim \bigg(\frac{m_3}{3m_1}\bigg)^{1/3}R_*\,.
\end{equation}
The cross-section for these TDEs can be obtained via
\begin{equation}
    \sigma_{\rm TDE} = \int_{\sum L_{\rm TDE}} db\,,
\end{equation}
where $\sum L_{\rm TDE}$ encompasses all the regions in the parameter space of the impact parameter $b$ that give a post-scattered star orbit with pericenter  smaller than $R_{\rm TDE}$. 

Solving for the post-scattered pericenter $r_{p,1}=a_1(1-e_1)=R_{\rm TDE}$ analytically is not straightforward. However, as shown in Section~\ref{sec:tde}, it is relatively easy to solve for $r_{p,1}=0$.
These solutions indicate the most probable impact parameter for TDE to happen. The TDE cross-section can be found by solving  $r_{p,1}=a_1(1-e_1)=R_{\rm TDE}$. But we have not been able to find an analytical solution for it. However, we can find a scaling for the TDE cross-section. That is $\sigma_{\rm TDE}\propto \sqrt{r}$.

Figure~\ref{fig:tde} shows the SMBH TDE cross-section as a function of the distance $r$ from the SMBH. These TDEs are completely forbidden in the inner region of the AGN disk in our setup since they require a high eccentricity of the stellar orbit. However, this is 
 almost impossible to obtain from the scattering between a 1$M_\odot$ star and a 30 $M_\odot$ BH, as evinced by Equation~\ref{eq:v1p}-\ref{eq:v2p} and shown in Figure~\ref{fig:SMBH}. SMBH  TDEs start to emerge around $10^4$ $r_g$ where the post-scattered eccentricity of the star orbit could reach unity. Type II and type III scatterings contribute most of the TDEs.

\subsection{Free-floating and high eccentricity planets}
For planetary systems with multiple planets, the interactions between planets may lead to chaotic evolution of the planet orbits, causing planet orbits to cross. Once the planet orbits can cross each other, the scattering between planets can significantly change the architecture of the planetary system. This scattering process will last until the two planet orbits become well separated or one of the planets is ejected from the system by a very close encounter \citep{Chatterjee08,Li14,Pu21,Li21}. In the latter scenario, the leftover planet is usually associated with high eccentricity \citep{Lin97,Ford08,Juric08,Spurzem09,Wang20,Li2020} and the ejected planet becomes a free-floating planet \citep{Sumi11, Beauge12} unless it is re-captured by other planetary systems. This is one of the potential mechanisms that can be used to explain the high eccentricity of exoplanets and free-floating planets.

Similar to the star-BH ejection case, the cross-section of the planet ejection is well described by Equation~\ref{eq:ej-be} to \ref{eq:ej_cross2}.
\begin{figure*}
\includegraphics[width=.95\columnwidth]{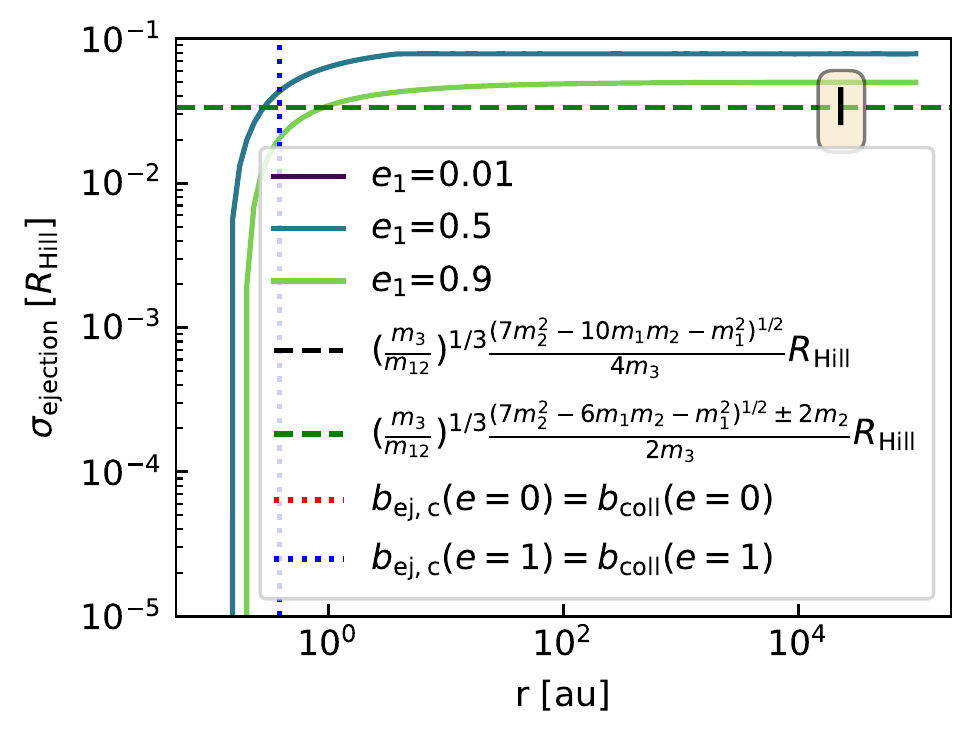}
\includegraphics[width=.95\columnwidth]{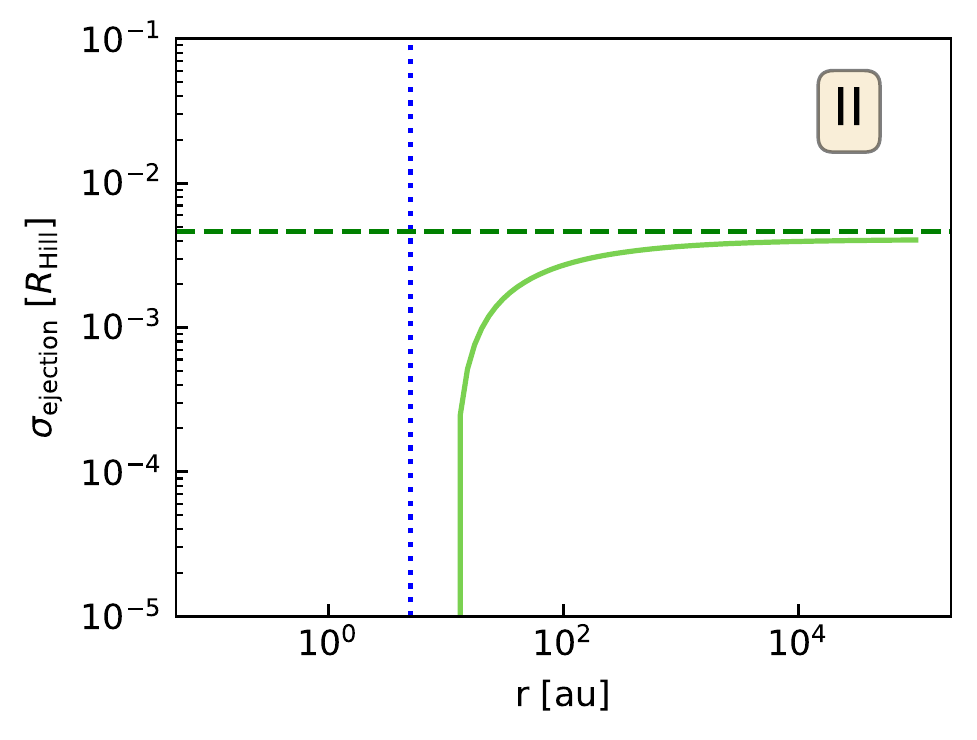}\\
\includegraphics[width=.95\columnwidth]{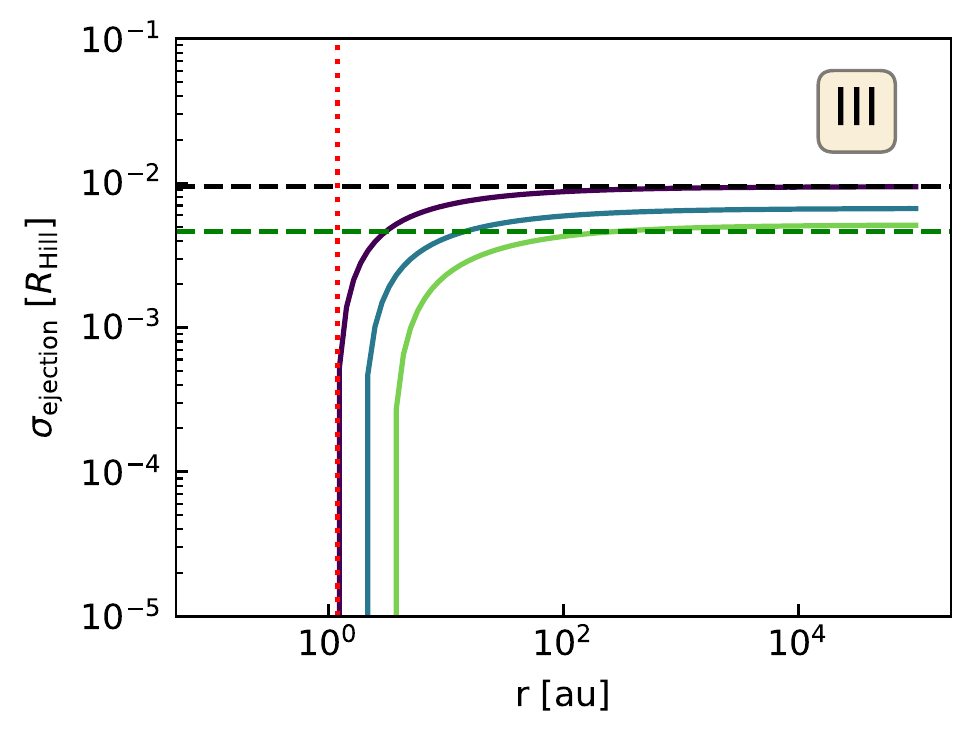}
\includegraphics[width=.95\columnwidth]{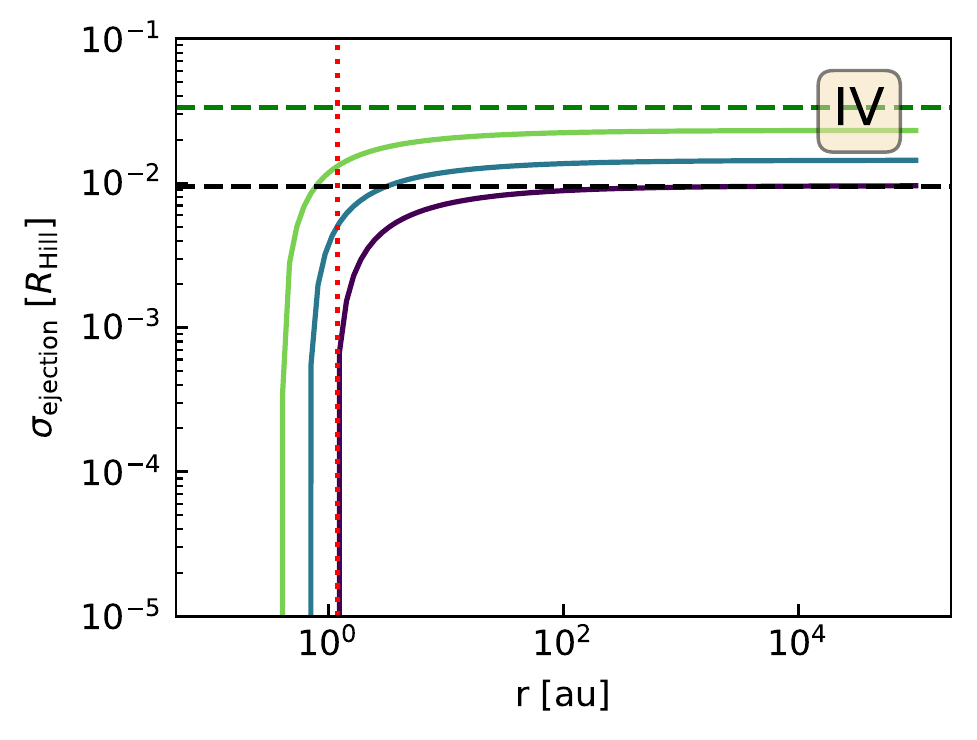}
\caption{Similar to Figure~\ref{fig:ejecction}, but the setup is an Earth-mass planet encounter with a Jupiter-mass planet under the potential of a solar mass star.}
\label{fig:p-ejection}
\end{figure*}
Figure~\ref{fig:p-ejection} shows the cross-section of the planet ejection between an Earth-mass planet and a Jupiter-mass planet. For prograde scatterings, it is difficult to get planet ejection from circular planet orbits, only high eccentricity orbits have the chance to be ejected from the system. The corresponding impact parameters between the two planets are roughly $10^{-3}R_{\rm Hill}-10^{-1}R_{\rm Hill}$. For retrograde scatterings, even circular planet orbits have the chance to be ejected, although retrograde scatterings are much rare in planetary systems. Figure~\ref{fig:p-ejection} also indicates that the ejections happen in the outer region of the planetary system. If the scatterings are too close to the host star, planet-planet collisions will occur.

Although Figure~\ref{fig:planet-ae} shows  equal mass, and equal orbital energy scatterings (ejection never happens), we can see from the trend if we compare it with Figure~\ref{fig:SMBH} that the leftover planet will become very eccentric once the other planet gets ejected from the system. This agrees with the general picture of planet-planet scattering in the literature.
\section{Conclusions}

\subsection{Summary}
We have presented a fully analytical solution to the 2-body scattering problem in the presence of a central gravitational potential, in planar geometry.
Our solution is highly accurate under the condition that the duration of the scattering event (as measured by the turning time of the scatterer's orbits) is much smaller than the orbital time in the potential of the third body. The valid parameter space in the circular scattering limit is given by Equation~\ref{eq:valid}, where for prograde scatterings, the impact parameter between two small objects needs to be smaller than the (well-known) Hill's radius,  and for retrograde scatterings, the impact parameter needs to be an even smaller number $\sim$ $(m_{12}/m_3)^{1/6}R_{\rm Hill}$.  For other scatterings with non-zero eccentricities, the critical impact parameter $b$ is between these two extreme values.

We tested the validity of our analytical solution via scattering experiments, and, as illustrative examples,  we applied the analytical solution to compute cross sections of astrophysical events in planar geometry under the presence of a third massive body providing the central potential.
These include scattering events between stars and black holes in the SMBH potential of AGN disks, which lead to micro-TDEs, where the star is disrupted by the stellar BH it is scattering with, and SMBH TDEs, where the scattered star ends up on a very eccentric orbit that leads it to a plunge within the tidal disruption radius of the SMBH, yielding an AGN-TDE. An accurate cross-section of these events is provided in Section~\ref{sec:app}. More generally, our analytical formulation can be very effective at saving computational time, while obtaining accurate results, in planar 2-body scatterings under the presence of a massive third body. From the calculated critical impact parameters and cross-sections, we can summarize some interesting results (Due to the geometry of the 2-D scattering setups, all cross-sections listed below are in length units). 

\begin{itemize}
    \item Micro TDEs in AGN disks are most contributed from prograde scatterings between a star and a stellar-mass black hole in low eccentricity orbits (hundreds to thousands of geometry cross-section of the tidal radius). For retrograde scatterings, if the scattering potion is $< 10^4 r_g$, the cross-section of the micro-TDE is basically the geometry cross-section of the tidal radius.
    \item The cross-section of star ejection (by a stellar-mass black hole) in AGN disks is roughly $10^{-5}-10^{-4}R_{\rm Hill}$, effectively independent of the scattering position with respect to the SMBH, although, in the inner region of the disk, the star gets tidally disrupted instead.
    \item Tidal disruption events by SMBH are relatively rare compared to micro-TDE in AGN disks. The cross-section is roughly $10^{-7}-10^{-4}R_{\rm Hill}$ for orbits without extreme initial eccentricity. The cross-section scales with scattering position in $r^{1/2}$. The inner region of the disk is nearly forbidden for macro-TDEs due to the micro-TDEs. 
    \item For planet-planet ejection, equal mass scatterings are more difficult to produce ejected planets than unequal mass scatterings. The cross-section for planet ejection is roughly $10^{-2}-10^{-1}R_{\rm Hill}$. The inner region is also forbidden for planet ejection due to the planet-planet collision.
\end{itemize}

\subsection{Caveats}
We emphasize that the analytical approximation we derived in the paper is meant to describe the two-body {\em single} scattering around a massive object. However, for prograde circular scatterings, where the two small objects orbit around the massive object in the same direction, multiple continuous scatterings can occur. In the scattering parameter space i.e. the space of impact parameter and relative velocity, these multiple scatterings can be hard to distinguish, e.g. the orbital turning from one scattering is not finished when the following scattering begins. The equations we derived require every single scattering to be finished, i.e. the turning process needs to be complete. Thus, Equation~\ref{eq:v1p} to \ref{eq:v2p} cannot properly describe close multiple scatterings for which each individual scattering cannot be discriminated. However, via numerical experiments, one can identify the regions of the parameter space where resonant scattering (multiple scatterings) does not occur, so that the equations derived here can be properly used.

For the environments of AGN disks,  two-body scatterings can be complicated due to the existence of the gaseous environment. These gas effects may lead to very different post-scattered results, thus significantly changing the cross-section/rate of the events we discussed in this paper. The descriptions and cross-section/rate estimates of the various astrophysical events made in this paper are all based on an assumption of no gas effects. More sophisticated computations, inclusive of gas effects, need to be performed to give more accurate cross-sections/rates of the various events. 

All calculations after Sec~\ref{sec:rmin} are based on the assumption of equal energy orbit scatterings. This is the most probable situation for scatterings between two Keplerian orbits. For more general cases, another free parameter indicating the energy ratio between the two planet orbits is required in all the expressions of the calculated critical impact parameters and cross-sections. 
\section*{Acknowledgements}
BM and KESF are supported by NSF AST 1831415 and Simons Foundation grant 533845. RP  acknowledges support by NSF award AST-2006839. ZZ acknowledges support by NASA award 80NSSC22K1413. YW thanks Scott Tremaine for reminding us of the work of Opik in the 1950s, and Dong Lai for helpful discussion. YW and BZ are supported by Nevada Center for Astrophysics.

\section*{Data Availability}
The data underlying this article will be shared on reasonable request to the corresponding author.



\bibliographystyle{mnras}
\bibliography{ref} 



\bsp	
\label{lastpage}
\end{document}